\documentclass[letterpaper,twocolumn,english,pre,showpacs,twocolumn,amsmath,amssymb,english]{revtex4}
\usepackage[utf8]{luainputenc}
\usepackage{color}
\usepackage{graphicx}

\makeatletter

\pdfpageheight\paperheight
\pdfpagewidth\paperwidth

\@ifundefined{textcolor}{}
{%
 \definecolor{BLACK}{gray}{0}
 \definecolor{WHITE}{gray}{1}
 \definecolor{RED}{rgb}{1,0,0}
 \definecolor{GREEN}{rgb}{0,1,0}
 \definecolor{BLUE}{rgb}{0,0,1}
 \definecolor{CYAN}{cmyk}{1,0,0,0}
 \definecolor{MAGENTA}{cmyk}{0,1,0,0}
 \definecolor{YELLOW}{cmyk}{0,0,1,0}
}

\usepackage[nooneline]{subfigure}
\usepackage{dcolumn}
\usepackage{bm}
\usepackage{babel}

\usepackage{comment}

\usepackage{babel}

\usepackage{babel}

\usepackage{babel}

\makeatother

\usepackage{babel}
\begin{document}

\title{The Effects of Front-Surface Target Structures on Properties of Relativistic
Laser-Plasma Electrons}

\author{S. Jiang, A.G. Krygier, D.W. Schumacher, K.U. Akli, R.R. Freeman }

\affiliation{Physics Department, The Ohio State University, Columbus, OH, 43210,
USA}
\begin{abstract}
We report the results of a study of the role of prescribed geometrical
structures on the front of a target in determining the energy and
spatial distribution of relativistic laser-plasma electrons. Our 3D
PIC simulation studies apply to short-pulse, high intensity laser
pulses, and indicate that a judicious choice of target front-surface
geometry provides the realistic possibility of greatly enhancing the
yield of high energy electrons, while simultaneously confining the
emission to narrow ($<5^{\circ}$) angular cones. 
\end{abstract}

\pacs{52.65.Rr,52.38.Kd,52.38.-r}

\maketitle

\section{Generation of Hot Electrons}

\subsection{Background}

The interaction between high intensity, short pulse lasers and solid
targets has been a subject of significant research over the last 15
years. The reason for this interest is the experimental observation
(and successful computer simulation) of the emission of ``hot electrons'',
that is, electrons that have energies far in excess of that corresponding
to emission from a thermal source. This hot electron production has
become an important research topic not only because it represents
a useful source of high energy electrons itself, but because it is
the mechanism for many important subsequent processes, including X
ray \textcolor{blue}{\cite{Murnane:1989}}, $\gamma$ ray \textcolor{blue}{\cite{Kmetec:1992,Schnurer:1995}},
and positron production \textcolor{blue}{\cite{Shearer:1973,Chen:PRL2009}},
and seeding even further processes such as ion acceleration \textcolor{blue}{\cite{Clark:2000}}.
Consequently there have been numerous efforts to increase the number
and energy of the hot electrons as well as to increase the efficiency of the conversion of laser energy
into hot electron energy.
Most of these studies have emphasized the role of the laser pulse
energy, duration, and intensity \textcolor{blue}{\cite{Wilks:1992,Beg:1997,Haines:2009,KlugeEnergy:2011}};
in addition, there is a robust literature describing the effect on
electron energy of a ``pre-plasma'' on the front of the target \textcolor{blue}{\cite{Paradkar:2011,Scott:2012,Ovchinnikov:2013}}.

\subsection{Target Design}

This paper discusses another means of enhancing the energy and directionality
of the laser generated hot electrons, quite apart from using higher
energy, more intense lasers. We propose, and verify through 3D PIC
simulation, target designs that include specific modifications of
the front surface geometry that not only substantially increase the
number of relativistic electrons from a target, but can greatly decrease
the solid angle of emission, all while maintaining or improving the
conversion efficiency of overall laser to electron energy.

There is a substantial body of experimental work on the effect of
surface roughness of targets on laser-plasma coupling, including the
laser absorption and the production efficiency and energy spectrum
of the generated x-rays or ions \textcolor{blue}{\cite{Palchan:2007,Rajeev:2003,Sumeruk:2007,Kulcsar:2000,Murnane:1993,Kahaly:PRL2008,Hu:POP2010,Margarone:PRL2012}.}
Various simulation works attribute the observed improvement to a surface
area increase and the local field enhancement introduced by the roughness \textcolor{blue}{\cite{Kupersztych:POP2004,Klimo:NJP2011,Andreev:POP2011,Kemp:POP2013}}.
Recently, several authors have shown that larger scale structures
can give rise to enhanced production of hot electrons and/or high
energy ions \textcolor{blue}{\cite{Kluge:2012,Gaillard:POP2011,Zigler:2013}}.
Kluge and Gaillard shot microcone targets on the cone walls and observed
a significant increase in electron and proton energies which they
attributed to direct laser acceleration of electrons along the cone
walls \textcolor{blue}{\cite{Kluge:2012,Gaillard:POP2011}}. Zheng
et al. proposed and simulated a ``slice cone'' target \textcolor{blue}{\cite{Zheng:POP2011}}
attributing the accelerated electrons to a similar mechanism. 2D simulations
on similar shaped nanobrush targets have also been published \textcolor{blue}{\cite{Cao:POP2010_1,Cao:POP2010_2,Zhao:POP2010,Cao:POP2011,Wang:2012,Yu:POP2012,Yu:APL2012}}.

Our target design employs a similar acceleration mechanism to that
of the previous work on microcones, but changes the structures to
periodic regular arrays. This change has the virtue of a target for
which the analysis and PIC simulations are easier to understand. Additionally,
this target configuration has the added feature of being easier to
align, in principle being no more difficult than a standard flat foil.
In the simulations described here we find that hot electrons start
off being extracted from the structures in a manner not dissimilar
to that reported by some previous 2D simulations \textcolor{blue}{\cite{Kluge:2012,Gaillard:POP2011,Baton:POP2008,Psikal:POP2010,Micheau:POP2010}}. However,
as we discuss in Sec. \textcolor{blue}{\ref{sec:2Dvs3D}}, fully 3D
simulations are required to reveal accurately the trajectories of
hot electrons.

\subsection{General Characteristics}

\begin{figure}[pht]
\centering{}\includegraphics[width=80mm]{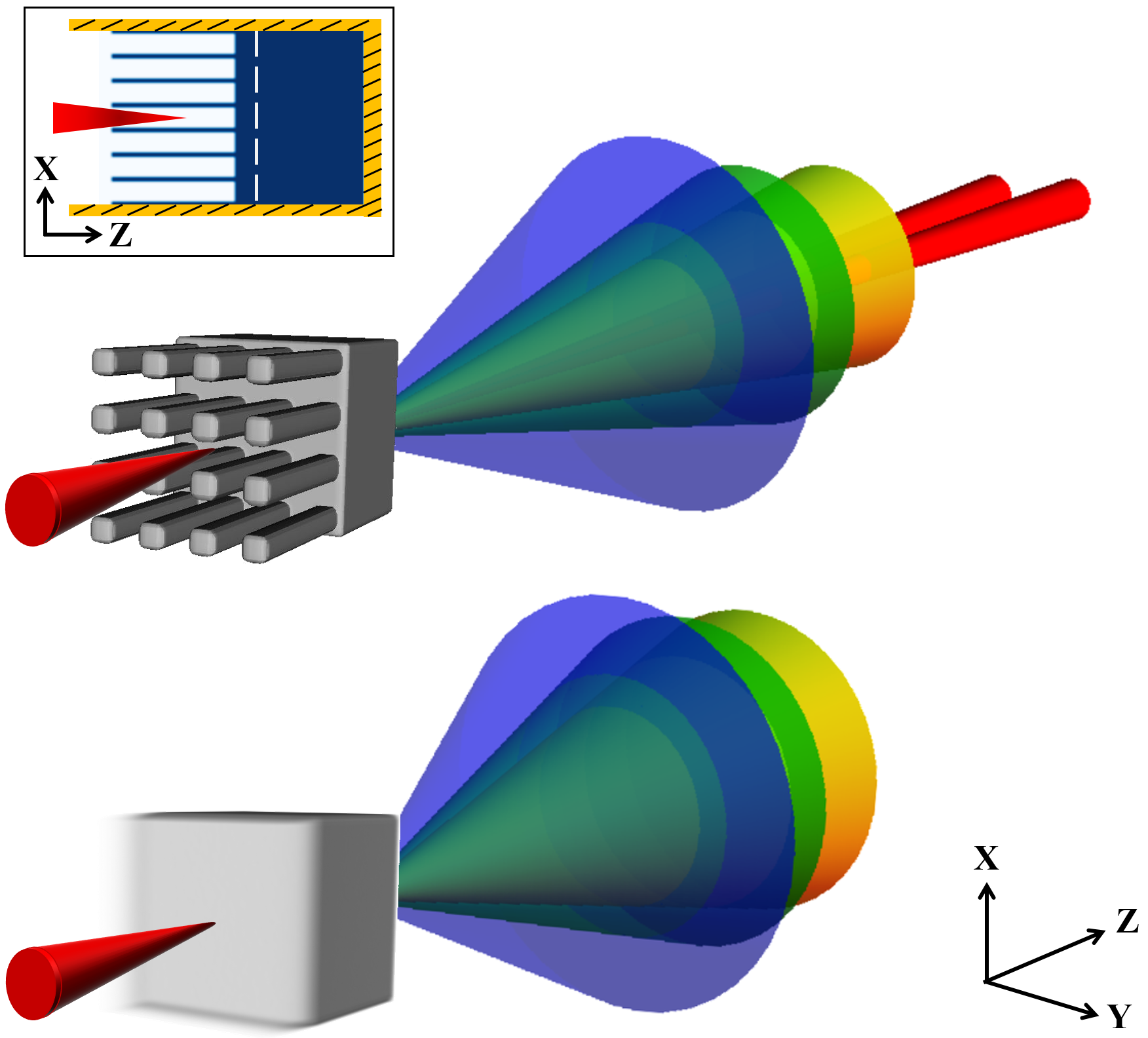} \caption{\label{fig:intro_cartoon} \emph{Schematic diagram of one target structure
investigated and a flat target with corresponding fast electron distributions.
The colored cones on the backside of the target represent the angular
distributions of electrons with different energies. Blue, green, yellow
and red cones indicate energies from low to high. (For example, for target and laser parameters we describe in \textcolor{blue}{\ref{sec:Our-simulations-use}}, the four colors indicate $1-10 MeV$, $50-60 MeV$, $80-100 MeV$ and $>150 MeV$ respectively.) The figure in the
upper left is the simulation setup for a 2D simulation; The shaded
yellow outer contour is modeled as a conductor.}}
\end{figure}

We use 3D PIC simulations to study the laser absorption and electron
spectrum changes due to front surface target structure. Fig. \textcolor{blue}{\ref{fig:intro_cartoon}}
is a schematic showing a general overview of the results from a 3D
PIC simulation of a specific geometric shape placed on the front of
a target compared to a regular flat target with pre-plasma. We have
chosen these shapes not only to facilitate the discussion of the electron
acceleration, but to highlight the limitations of 2D PIC simulations
as well. The colored cones on the backside of the target represent
the angular distributions of electrons with different energies. Blue,
green, yellow and red cones indicate electron energies from low to
high. For a regular flat target with pre-plasma (shown at the bottom),
electrons usually have divergence angles between $30-60^{\circ}$ \textcolor{blue}{\cite{Stephens:PRE2004,Akli:PRE2012}}.

For a target with tower structures on the front (shown at the top),
there are considerably more higher energy electrons generated. The
highest energy hot electrons form into two narrow cones lining in
the y direction (red), whereas the laser is polarized along x. The
inset on the top left shows a typical 2D simulation setup, where only
x and z dimensions can be modeled.

\section{Simulation Setup}

\begin{figure}[pht]
\centering{}\includegraphics[width=80mm]{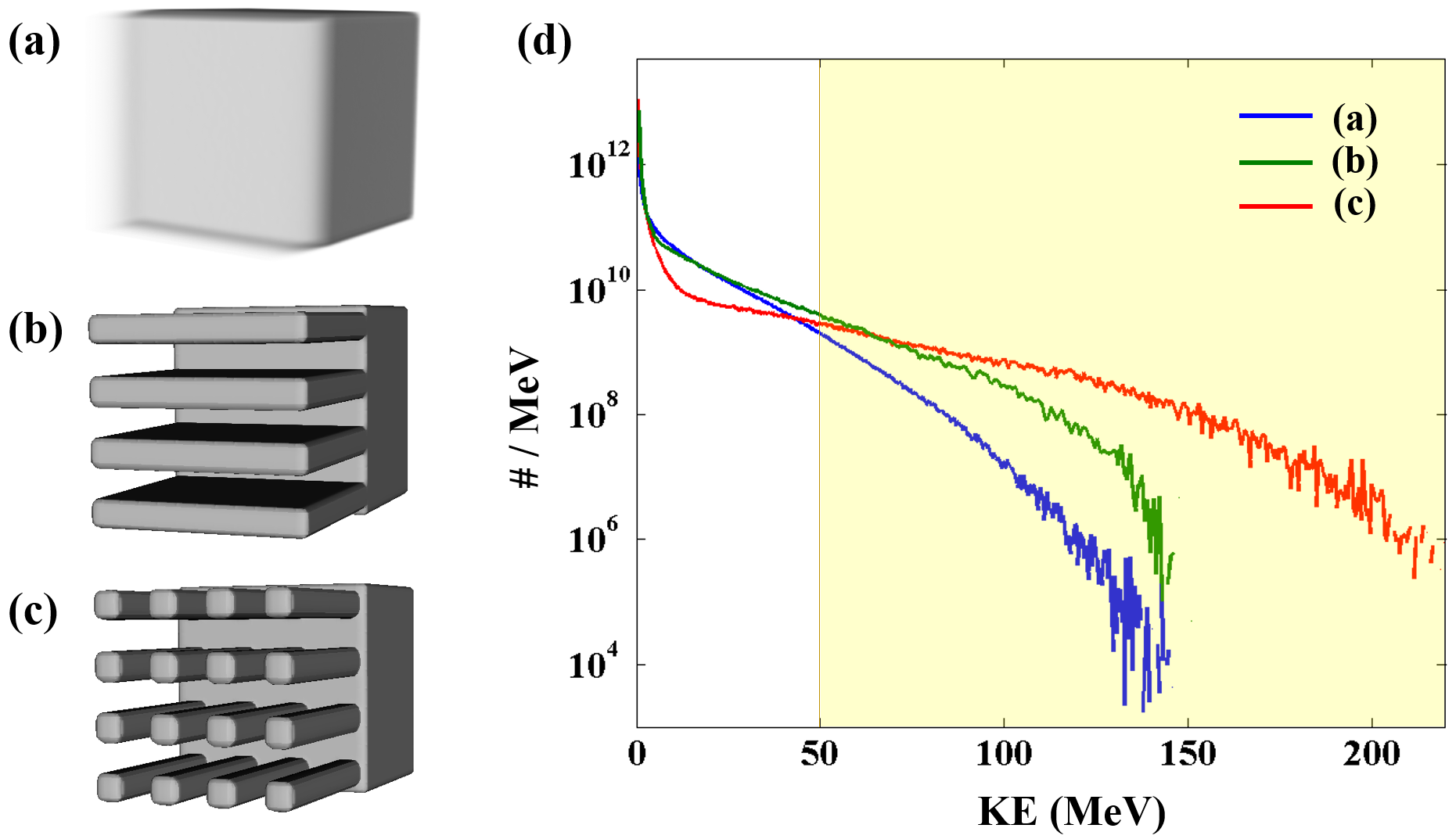} \caption{\label{fig:3struct_fE} \emph{Electron spectrum for different targets
from 3D simulations. There are 3 types of targets: (}\textcolor{blue}{\emph{a}}\emph{)
a flat target with $1\mu m$ pre-plasma; (}\textcolor{blue}{\emph{b}}\emph{)
a target with slab structures on the front, the depth and width of
the slabs are $10\mu m$ and $1\mu m$. The spacing between the slabs
is $2\mu m$; (}\textcolor{blue}{\emph{c}}\emph{) a target with tower
structures on the front, the depth and width of the towers are also $10\mu m$ and $1\mu m$. The spacing between towers in both transverse directions
is $2\mu m$. (}\textcolor{blue}{\emph{d}}\emph{) shows the electron
energy spectra of targets (}\textcolor{blue}{\emph{a}}\emph{), (}\textcolor{blue}{\emph{b}}\emph{)
and (}\textcolor{blue}{\emph{c}}\emph{). The blue curve for target
(}\textcolor{blue}{\emph{a}}\emph{) is commonly observed in experiments
on flat targets. The green and red curves for targets (}\textcolor{blue}{\emph{b}}\emph{) and (}\textcolor{blue}{\emph{c}}\emph{) show substantial increases in the yield of electrons above $50 MeV$.}}
\end{figure}

Our simulations use the 3D PIC code LSP \textcolor{blue}{\cite{Welch:2006}}.
Three types of targets are studied, and are indicated by Fig. \textcolor{blue}{\ref{fig:3struct_fE}(a)},
\textcolor{blue}{(b)} and \textcolor{blue}{(c)}. 

\label{sec:Our-simulations-use}Fig. \textcolor{blue}{\ref{fig:3struct_fE}(a)}
shows a normally flat target with an exponentially decaying pre-plasma
on the front. The scale length of the pre-plasma is $1\mu m$. The
target in Fig. \textcolor{blue}{\ref{fig:3struct_fE}(b)} is chosen to be an example of
a target that apparently lends itself to a 2D simulation; we refer
to this as a slab target. The slabs are $10\mu m$ deep in the laser
z direction, and $1\mu m$ wide in the perpendicular laser polarization
direction. The spacing between the slabs is $2\mu m$. The third type
is shown in\textcolor{blue}{(c)}, what we have labeled as a tower
target. The towers are also $10\mu m$ deep and $1\mu m$ wide. Here
again the transverse spacings are $2\mu m$. Both structures in \textcolor{blue}{(b)}
and \textcolor{blue}{(c)} have sharp interfaces. The base of each
target is $11\mu m$ on a side. The material is Al, initialized as
singly ionized, but subsequently treated by ADK ionization model;
collisions are not included. (2D simulations using Spitzer cross-sections with collision frequency capped to a maximum value of $2\times10^{16} s^{-1}$ do not show significant changes in hot electron distributions either spectrally or angularly.) The simulation box is made of $120\times120\times600$
cells with mesh sizes $\Delta x=\Delta y=0.1\mu m$, $\Delta z=0.05\mu m$.
(We checked the appropriateness of these parameters using short simulations
with $120\times120\times1200$ cells and $\Delta z=0.025\mu m$. We
found little difference in either the energies or the trajectories
of the hot electrons compared to the coarser grid.) The boundary conditions
are absorbing, and the time step is 0.03fs. In the simulation, a Gaussian
laser pulse is used with a wavelength of $800nm$, $15J$ energy,
$30fs$ FWHM pulse duration and a focal spot diameter of $2.9\mu m$,
yielding $5\times10^{21}W/cm^{2}$ peak intensity. Electron energy
and spatial distributions are measured at a plane $5\mu m$ inside
the target. In all of our simulations reported here, we do not take
into account changes to the electron spectrum due to target charging
when the hot electrons leave the target, but we note that charging
will have a minimal effect on the simulation for the higher energy
($>100$ MeV) hot electrons that are of interest here \textcolor{blue}{\cite{Link:POP2011}}.
Finally, we employed a direct-implicit advance with an energy conserving
particle push which greatly reduces numerical heating.

\section{2D vs. 3D Simulations}

\label{sec:2Dvs3D}

While we have found it important to use full 3D PIC simulations in
our study, the use of 3D simulations is not common in the literature
of hot electron production. Largely because of the often prohibitively
large computational demands of 3D PIC simulations, 2D PIC simulations
of hot electron production have been the de facto standard. However,
in our studies we observe multiple features of the hot electrons'
behavior in 3D that simply cannot be addressed in 2D. For example,
note that our slab (Fig. \textcolor{blue}{\ref{fig:3struct_fE}(b)})
and tower structures (Fig. \textcolor{blue}{\ref{fig:3struct_fE}(c)})
\emph{cannot be differentiated} using a 2D simulation; while in fact,
we find that using 3D simulations the electron acceleration dynamics
are significantly different for the two geometries. These differences
show up most clearly and significantly in the predicted angular spectrum.
It is only with a fully 3D simulation that we discover the remarkably
narrow angular divergence pattern discussed in \textcolor{blue}{\ref{angular distribution}}.

The use of 2D results as a proxy to predict 3D electron energy spectra
and angular distributions requires significant and in many cases suspect
assumptions. Fundamentally, the conversion is only possible when the
desired quantity depends only on aspects of the target-laser interaction
such as material, degree of pre-pulse etc. that have no transverse
asymmetry. For a linearly polarized laser pulse striking a plane,
unstructured target, the low energy part of the 2D simulated energy
spectrum of hot electrons can be generalized to 3D by appropriately
symmetrizing the 2D spectrum through 360 degrees. This method can
be useful for calculating quantities that are dominated by low energy
electrons, such as $K_{\alpha}$ radiation, but not simulating the
high energy electrons in the kind of structured targets we discuss
here.

In this study our main interest is in the highest energy electrons.
As we show later, these are produced by direct laser acceleration.
The inset in the upper left of Fig. \textcolor{blue}{\ref{fig:intro_cartoon}}
shows a setup for both towers and slabs if 2D simulations were to
performed. Clearly this configuration fails the test for meaningful
2D simulation, for the target geometry is not cylindrically symmetric
within the laser spot dimensions and the laser-plasma dynamics clearly
do not depend only on local conditions. There are other, more subtle
reasons why this specific problem cannot be treated in 2D: The background
plasma density generated by the short pulse laser itself is higher
in 2D than in 3D because of the translational invariance in the virtual-y
dimension. As we discuss below, the maximum electron energy from DLA
is quite sensitive to the background plasma density. In addition,
because the transverse dimensions of our structures are on the order
of the laser wavelength, they tend to act as waveguides which changes
the laser phase velocity. The DLA mechanism is strongly affected by
the phase velocity of the laser relative to the accelerating hot electron,
so that only a 3D simulation can capture this physics (see \textcolor{blue}{\ref{DLA}})
.

\section{3D Simulation Results}

\subsection{Energy Spectrum}

The electron energy spectrum for the 3 types of targets is shown in
Fig. \textcolor{blue}{\ref{fig:3struct_fE}(d)}. These simulations
are for the laser conditions outlined in \textcolor{blue}{\ref{sec:Our-simulations-use}}.
The blue curve is for the simple unstructured flat target. Pre-plasma,
modeled as an exponential with a scale length of $1\mu m$, is included
in all of our unstructured flat target simulations because most experimental
situations involve generation of some pre-plasma and because it provides
a more interesting comparison since pre-plasma increases the laser
absorption and coupling to the target. We calculate a $17.9\%$ conversion
efficiency from laser energy to fast electron kinetic energy (electrons
with energies $>1MeV$). For the slab type target \textcolor{blue}{(b)},
the conversion efficiency is enhanced to $23.0\%$, with substantially
more electrons generated above $50MeV$. The tower target yields yet
another, different spectrum. With a conversion efficiency of $16.6\%$,
comparable to that of the flat target with pre-plasma, the spectrum
shows a large reduction in the yield of low energy electrons and a
dramatic increase in the yield of high energy electrons. Specifically,
the conversion efficiencies for fast electrons with energies above
$50MeV$ are $1.5\%$, $5.2\%$ and $7.9\%$ for the flat, slab and tower targets respectively.
This is a significant re-shaping of the energy spectrum with respect
to that from a flat target such that the low energy portion of the
electron spectra is shifted to higher energies. These target geometries
substantially increase the number of electrons with the highest energies,
while maintaining the overall efficiency of coupling. This result
constitutes one of the two primary findings of our work. The other
concerns the electron angular distribution.

\subsection{Angular Distribution}

\label{angular distribution} 
\begin{figure}[pht]
\centering{}\includegraphics[width=80mm]{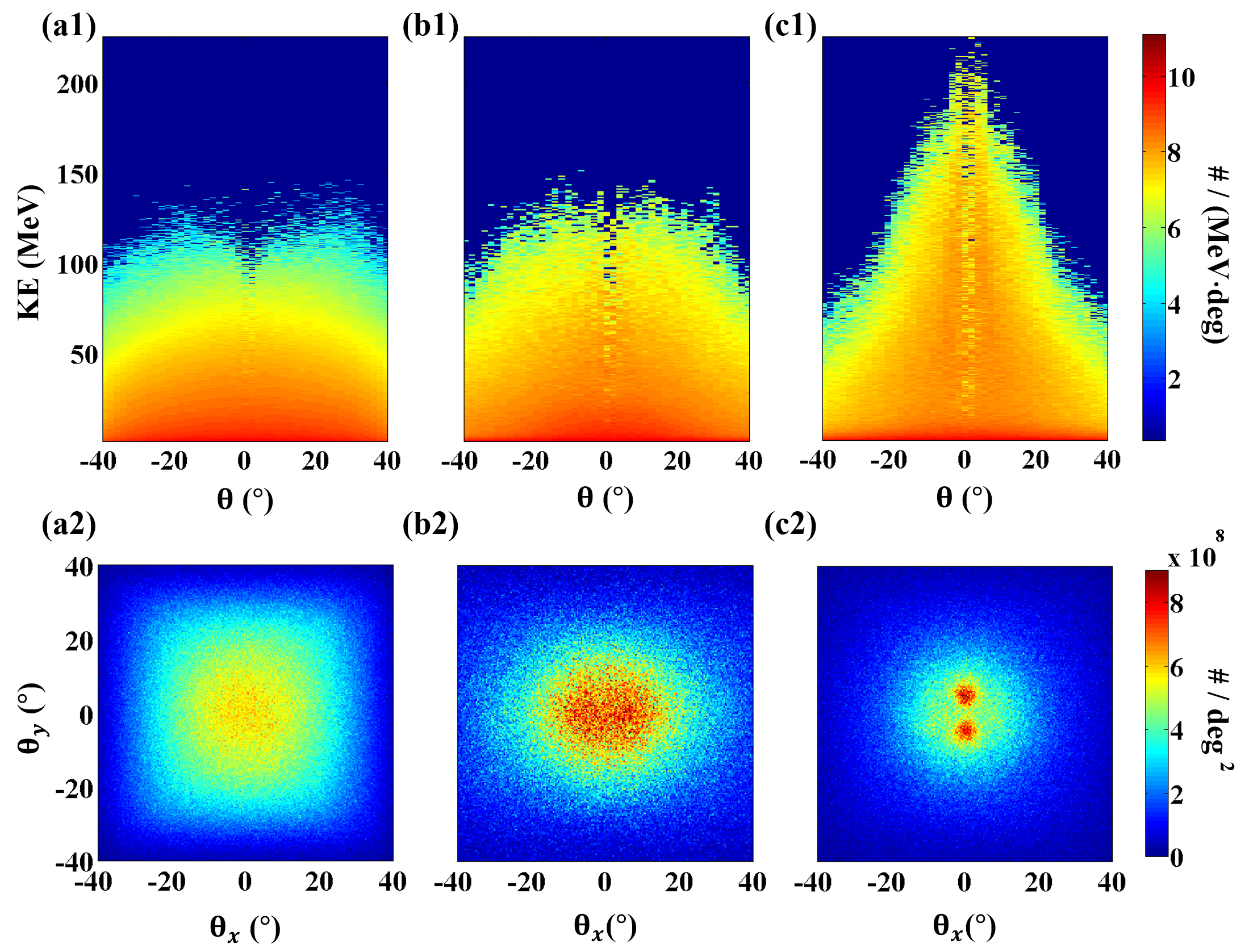} \caption{\label{fig:ang_distrib_3D} \emph{Angular distribution of fast electrons
( $>1MeV$). The top three graphs are fast electron number distributions
(on a color log scale) as a function of kinetic energy and angle $\theta$
with respect to the forward (z) direction. The bottom three graphs
are fast electron number distributions as a function of direction
in the xz and yz planes. (}\textcolor{blue}{\emph{a1}}\emph{), (}\textcolor{blue}{\emph{a2}}\emph{)
correspond to the flat target in Fig. }\textcolor{blue}{\emph{\ref{fig:3struct_fE}(a)}}\emph{.
An overall divergence angle of about $60^{\circ}$ is seen. (}\textcolor{blue}{\emph{b1}}\emph{),
(}\textcolor{blue}{\emph{b2}}\emph{) correspond to the slab-structured
target in Fig. }\textcolor{blue}{\emph{\ref{fig:3struct_fE}(b)}}\emph{.
The over all divergence angle is about $40^{\circ}$. (}\textcolor{blue}{\emph{c1}}\emph{),
(}\textcolor{blue}{\emph{c2}}\emph{) correspond to the tower-structured
target in Fig. }\textcolor{blue}{\emph{\ref{fig:3struct_fE}(c)}}\emph{.
The distribution shows two peaks at $\theta_{y}\approx\pm4-5^{\circ}$.
For electrons $>100MeV$, each cone angle is about $4-5^{\circ}$. }}
\end{figure}

The simulations show a striking modification
of the angular distribution by our targets, as is shown in Fig. \textcolor{blue}{\ref{fig:ang_distrib_3D}}.

The top row is electron number distribution as a function of kinetic
energy and angle. The angle indicates the direction of electron velocities
in a solid angle, $\Delta\Omega=2\pi sin\theta\Delta\theta$, where
$\theta=tan^{-1}(\frac{\sqrt{p_{x}^{2}+p_{y}^{2}}}{p_{z}})$. The
bottom row shows the 2D projected divergence map of fast electrons
($>1MeV$). The three graphs are the fast electron number distributions
as a function of $\theta_{x}$ and $\theta_{y}$, where $\theta_{x}=\pm tan^{-1}(\frac{p_{x}}{p_{z}})$,
$\theta_{y}=\pm tan^{-1}(\frac{p_{y}}{p_{z}})$. Subplots \textcolor{blue}{(a1,a2)},
\textcolor{blue}{(b1,b2)} and \textcolor{blue}{(c1,c2)} correspond
to targets \textcolor{blue}{(a)}, \textcolor{blue}{(b)} and \textcolor{blue}{(c)}
in Fig. \textcolor{blue}{\ref{fig:3struct_fE}} respectively. Comparing
\textcolor{blue}{(a1)}, \textcolor{blue}{(b1)} and \textcolor{blue}{(c1)},
in general the electron divergence reduces with increasing energy.
However, for the flat target, the decrease is small compared to the
two types of structured targets. While some collimation is seen in
the high energy portion of the electrons generated by the slab target
\textcolor{blue}{(b1)}, the tower target \textcolor{blue}{(c1)} shows
a significant improvement in the collimation. This result is clearer
in the bottom graphs of Fig. \textcolor{blue}{\ref{fig:ang_distrib_3D}}. The flat target
\textcolor{blue}{(a1)} shows a cylindrically symmetric angular distribution,
with a large divergence angle of about $60^{\circ}$. Both the shape
of slab and tower breaks the rotational symmetry, so that, as expected,
the corresponding angular distributions do not maintain such symmetry.
For the slab target \textcolor{blue}{(b1)} , the distribution is wider
in the x direction than in the y direction. For the tower target \textcolor{blue}{(c1)},
although the target shape itself has $90^{\circ}$ rotational symmetry,
the angular distribution does not. The distribution shows 2 peaks
centered at $\theta_{y}\approx\pm4-5^{\circ}$, each peak is about
$4-5^{\circ}$ FWHM.

Remarkably, in the case of the tower target, the highest energy electrons
tend to preferentially fall into these 2 small cones. For electrons
$>100MeV$, approximately $30\%$ of the electrons (or $5.7\times10^{9}$
in number) fall into the two $5^{\circ}$ full angle cones. We find
that for the tower targets the fast electron bunch has a pulse duration
of $42fs$. More importantly, for electrons $>100MeV$, the pulse
is even shorter, about $13fs$ leading to an average current of $70kA$.
Assuming a source diameter of approximately $3\mu m$, this electron
source brightness at $100MeV$ is on the order of $10^{23}s^{-1}mm^{-2}mrad^{-2}(0.1\%\ bandwidth)$.

\section{Discussion}

We attribute the enhancement in the high energy electron spectrum to
direct laser acceleration (DLA) \textcolor{blue}{\cite{JHYang:Thesis,Andy:Thesis,Andy:unpublished}}. Electrons undergoing DLA gain energy directly from
the laser fields. This mechanism requires the accelerated electrons
to be injected into the pulse at the right phase, and then to travel
along with the pulse over a significant length, and then be injected
into the target at the critical surface.

\subsection{Direct Laser Acceleration}

\label{DLA}

In DLA the highest energies are achieved by electrons which are optimally
injected: injected at a point as far from the critical surface as
allowed by the laser and evolving plasma and at an ideal phase of
the laser. By its nature, DLA is exquisitely sensitive to the initial
velocity and position of the accelerated electron relative to the
phase of the laser electric field. If a low energy electron is injected
into a weak plane wave at a node of the electric field, the electron
will gain energy in the first half of the laser cycle and then return
it to the pulse during the second half of the cycle. Net energy gain
is possible if the electron can remain in the accelerating half-cycle
of the laser field long enough and, crucially, can subsequently escape
before experiencing the decelerating half-cycle. When the laser intensity
is well into the relativistic regime, the Lorentz force accelerates
the electron's longitudinal velocity to roughly the speed of light
within the first quarter cycle of the wave. Provided the phase velocity
of the laser pulse is close to $c$ as well, the appropriately phased
moving electron will experience a positive acceleration from the wave
over an extended distance and will gain substantial energy. For example,
an electron starting at rest introduced into a node of a plane wave
in vacuum with an intensity of $5\times10^{21}W/cm^{2}$ will accelerate
for a distance of $272\mu m$, gaining a peak energy of $1.14GeV$
before it begins to decelerate. The purpose of the structure on
the front surface of the target is to provide a means for inserting
the electrons into the plane wave (see below) as well as permit an
extended acceleration length followed by an abrupt exit (at the relativistic
critical surface). It is important to note that the vast majority
of the electrons, even in the structured targets we propose, do not
fulfill the requirement of DLA \textcolor{blue}{\cite{Andy:Thesis,Andy:unpublished}}. This is consistent with the observation that only
a small proportion of the laser-plasma electrons are observed to have
high energy \textcolor{blue}{\cite{Link:POP2011}}.

In a plasma the phase velocity of the laser pulse depends upon the
plasma density and is given by $v_{ph}=c/\sqrt{1-(\frac{\omega}{\omega{}_{p}})^{2}}$
, where $\omega_{p}^{2}=\frac{n_{e}e^{2}}{m\epsilon_{0}}$; thus the
de-phasing length for an electron undergoing DLA is also sensitive
to plasma density. As an example, for an intensity of $5\times10^{21}W/cm^{2}$
, if $n_{e}=0.02n_{cr}$, where $n_{cr}$ is the critical density
($=\frac{m\epsilon_{o}\omega^{2}}{e^{2}})$, the acceleration length
is about $25\mu m$ and the maximum energy is $170MeV$; however,
increasing the density by a factor of 5, the acceleration length reduces
to $7\mu m$ with a maximum energy of $70MeV$. When we compare the
2D simulation results for the slabs with 3D simulations for both the slabs and the towers, we find the
electron cut-off energy for 3D towers to be much higher.
This is because the 3D simulation of the towers has the most vacuum space
thus lowest background electron density. With lower background density
the acceleration length is longer. As discussed in \textcolor{blue}{\ref{sec:2Dvs3D}},
this is one reason why 3D simulations are required to accurately predict
the production mechanism of hot electrons on non-flat targets. Although 3D slabs have a lower background density than 2D slabs, the cut-off energy is lower due to a different effect that we discuss in \textcolor{blue}{\ref{3D_confinement}}.

\subsection{Electron Injection}

While the DLA mechanism makes it possible in principle to accelerate
electrons to significant energies, the practical issue that arises
is the need for placing, or injecting, the electrons into the laser
field at the proper position and time relative to the oscillating
laser fields. A successful injection technique must fulfill four conditions:
1) A significant number of electrons need to be injected; 2) The position
where the electrons are injected needs to be far enough from (relativistic)
critical density so that there is a sufficient acceleration length;
3) The point of injection should be at a position where the laser intensity
is high; 4) Over the course of the acceleration length there must
be a channel with a relatively low electron
density so the laser can propagate with a phase velocity on the order
of, but not significantly larger than, $c$.

\begin{figure}[pht]
\centering{}\includegraphics[width=80mm]{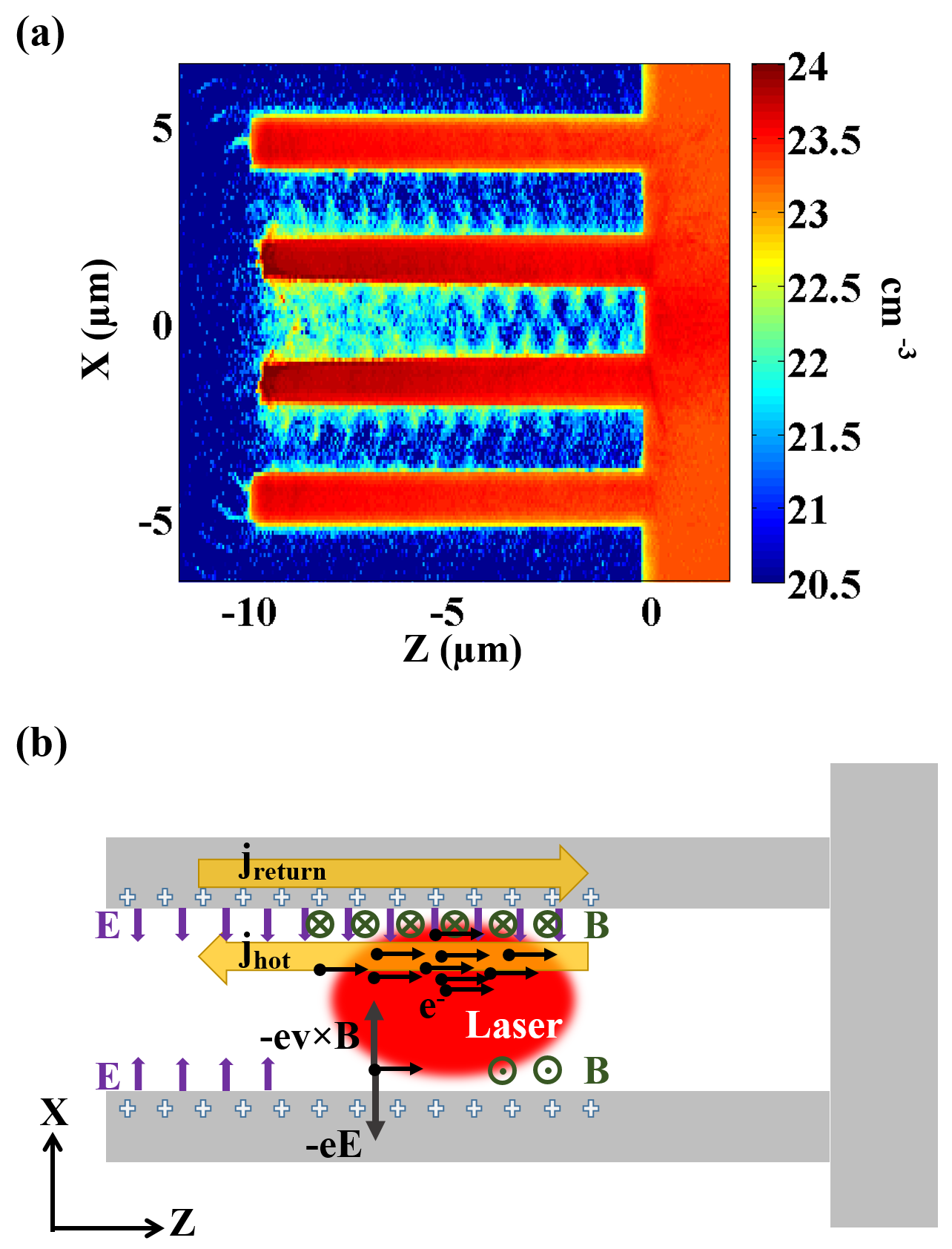}
\caption{\label{fig:physics_2D} \emph{Electron injection and acceleration
mechanisms as revealed in a 2D simulation. (}\textcolor{blue}{\emph{a}}\emph{)
is a plot of electron density on log scale when the peak of the laser
pulse is located at $Z=-4\mu m$. Electron bunches are pulled out
from the structure surfaces by the laser E field. (}\textcolor{blue}{\emph{b}}\emph{)
is a cartoon showing the accelerating and guiding mechanisms of the
electrons pulled out into the gaps. The red ellipse represents the
laser pulse. The black dots are the electrons. The trajectories shown
here are substantially different from those revealed using fully 3D
simulations (see Fig. \textcolor{blue}{\ref{fig:track_GRT_SPK}}).}}
\end{figure}

With these conditions in mind, we now address DLA using Fig. \textcolor{blue}{\ref{fig:physics_2D}}\textcolor{blue}{(a)},
which shows the electron density from a 2D simulation. (Below we compare
this 2D simulation to 3D next where we find that the injection process
is similar but the electron confinement to be very different). The
laser is incident from the left and the plot is at a time when the
laser pulse is roughly half way down the length of the structures.
The target and laser parameters are the same as in the 3D simulations:
the transverse gap size is $2\mu m$, the structures are $1\mu m$
in width, and $10\mu m$ in depth. The electrons are pulled out of
the structures by the laser E field, as is indicated by the bunches
along the structure surfaces. There are several noteworthy characteristic
features: The bunches are located in such a way that they are separated
by one laser wavelength on one side and are $\pi$-phase shifted on
opposite facing sides, corresponding to regions where $\vec{E}\cdot\hat{n}<0$
($\hat{n}$ is the surface normal). At a given time, the bunching
maximizes at the point of the electric field anti-nodes. For this
intensity, $5\times10^{21}W/cm^{2}$, each bunch has an electron density
of more than $10^{22}/cm^{3}$.The electron density in the center
of the gaps is considerably lower such that the pulse has a phase
velocity approximately equal to the speed of light. These bunches
constitute the injected electrons into the DLA acceleration channel:
the final energy of any bunch depends upon when it is formed relative
to the laser pulse and the position of its original location. The
most energetic electrons originate from the tip of the structures,
as long as the structure is shorter than the de-phasing length.

We have found through trial and error that the maximum number of the
highest energy electrons occurs when the spacing between the structures
is slightly smaller than the laser focal spot, and the laser intensity
at the structure surface is sufficiently large to ionize electrons
while simultaneously not increasing the electron density in the acceleration
channel to the point that the acceleration lengths are reduced due
to increased phase velocity of the laser pulse. A full optimization
of our structures would require numerous 3D simulations, and these
have not been done here. In fact we know that the target size parameters
are certainly \emph{not} optimum: in order to keep the grid size small,
and the 3D simulation manageable, we chose a small focal spot so we
could have a small spacing between the structures. However, the problem
is not scalable: $2\mu m$ spacing between the structures results
in some wave-guiding of the $0.8\mu m$ wavelength laser (with its
attendant modification of the phase velocity) and the background density
between the structures fills too rapidly to preserve the acceleration
length.

\subsection{Electron Confinement in 2D}

\label{subsec:e_confine}

In our 2D simulations we observe that electrons are guided by the
structure shape. This observation is consistent with other recent
2D simulations \textcolor{blue}{\cite{Sentoku:POP2004,Nakamura:PRL2004}}.
The responsible physics based upon the 2D simulation is diagrammed
in Fig. \textcolor{blue}{\ref{fig:physics_2D}(b)}. Once electrons
are pulled out into the pulse, some of them with the right initial
conditions accelerate with the laser pulse. They form a current density
$j_{hot}$ near the slab surface, leaving positive charges at the
surface of the structures. These positive charges draw return currents
just inside the structure with a current density of $j_{return}$
in the opposite direction of $j_{hot}$. Now these two surface currents
are uniform in the virtual-y direction in a 2D xz coordinate system,
giving rise to a $B_{y}$ field also uniform in the y direction. At
the same time, the charge separation induces an electric field pointing
away from the structure surface. These two fields, $B_{y}$ and $E_{x}$,
can be quite large; in our simulations, they are approximately 0.3
of the peak laser B and E fields. When the electron velocity in the
z direction is approximately the speed of light the Coulomb force
from $E_{x}$ and the magnetic force from $cB_{y}$ are equal and
opposite along the surface of the slab, resulting in electrons being
guided forward along the structure.

\begin{figure}[pht]
\centering{}\includegraphics[width=80mm]{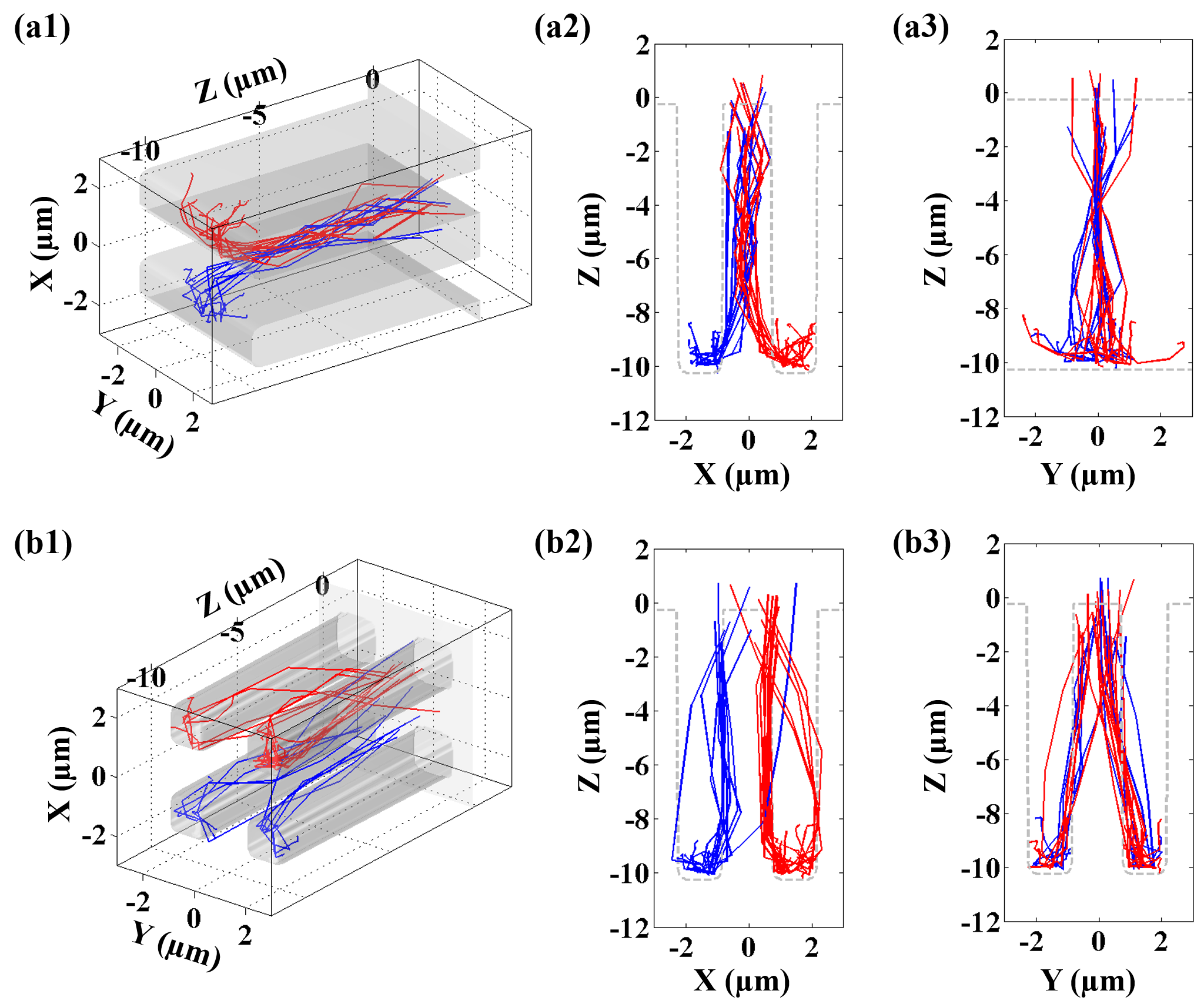} \caption{\label{fig:track_GRT_SPK} \emph{Trajectories of the higher-energy hot
electrons ($>120MeV$) from 3D simulations. (}\textcolor{blue}{\emph{a1}}\emph{),
(}\textcolor{blue}{\emph{a2}}\emph{), (}\textcolor{blue}{\emph{a3}}\emph{)
correspond to the slab target Fig. }\textcolor{blue}{\emph{\ref{fig:3struct_fE}(b)}}\emph{.
(}\textcolor{blue}{\emph{a2}}\emph{), (}\textcolor{blue}{\emph{a3}}\emph{)
are side-views while looking along either y-axis or x-axis. Red curves
indicate electrons with initial positions $X>0$, while blue are those
from $X<0$. In the case of slabs, electrons bounce back and forth
between the structure surfaces in the x direction, and are bent towards
the center in the y direction. (}\textcolor{blue}{\emph{b1}}\emph{),
(}\textcolor{blue}{\emph{b2}}\emph{), (}\textcolor{blue}{\emph{b3}}\emph{)
correspond to the tower target Fig. }\textcolor{blue}{\emph{\ref{fig:3struct_fE}(c)}}\emph{.
Electrons are guided by the structure surfaces in the x direction,
and pinch in the y direction. These trajectories are considerably
different from those predicted using 2D simulations (see Fig. \textcolor{blue}{\ref{fig:physics_2D}})}}
\end{figure}

\begin{figure}[pht]
\centering{}\includegraphics[width=80mm]{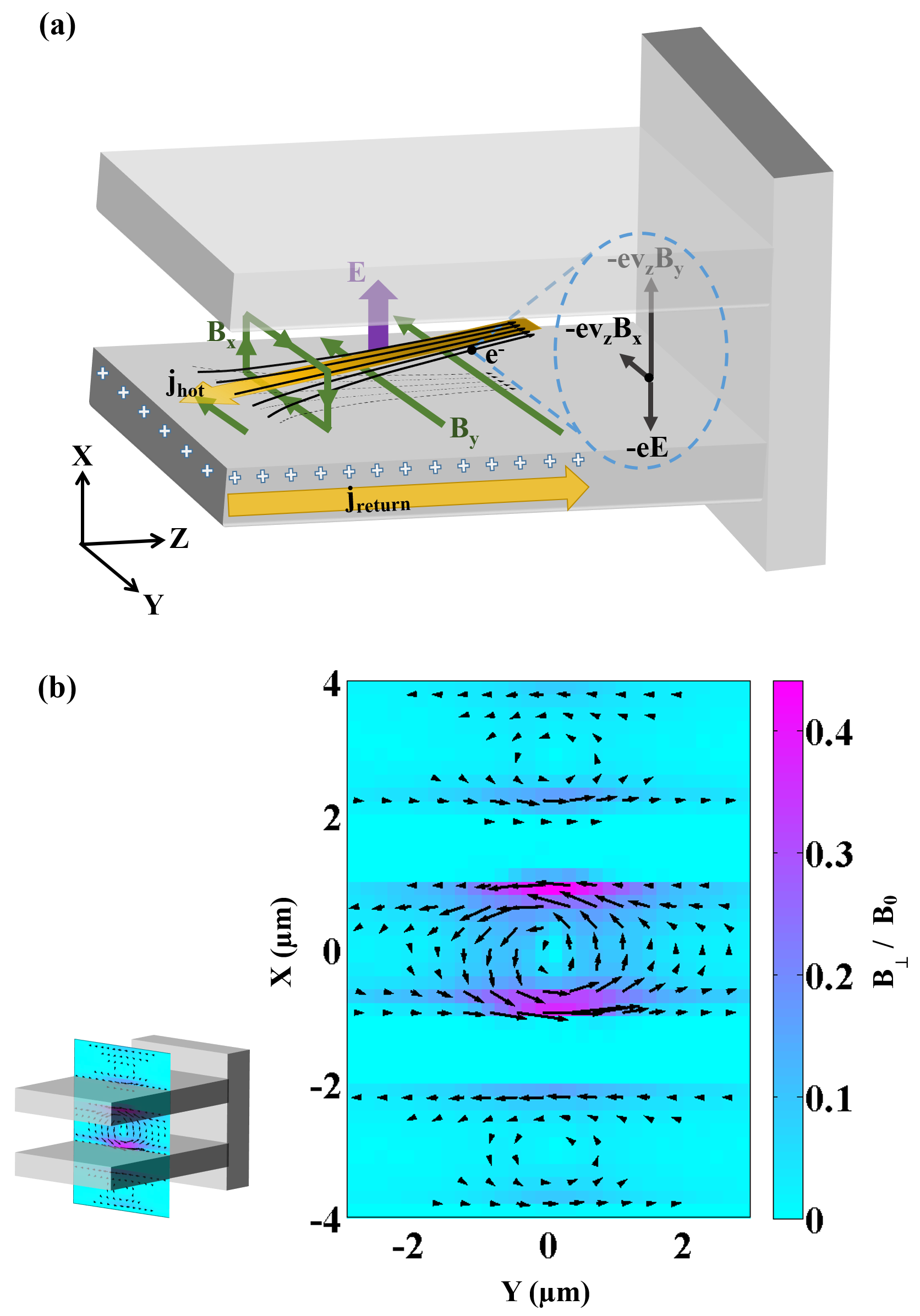} \caption{\label{fig:physics_fin} \emph{Guiding mechanisms in 3D slabs. In
(}\textcolor{blue}{\emph{a}}\emph{), in the x-direction, electrons
are bounced back and forth between surfaces, while in the y-direction,
they are guided towards the center. (}\textcolor{blue}{\emph{b}}\emph{)
is a plot of the B field perpendicular to the laser direction at $Z=-5\mu m$,
where the peak of the laser pulse is located. A spatial average over
$\lambda$ in z direction is applied to minimize the oscillating $B_{y}$
field from the laser. $B_{\perp}$ is normalized by the peak $B$
field of the laser $B_{0}$. The color scheme indicates the magnitude
while the arrows point out the directions. The lengths of the arrows
are proportional to the magnitude of $B_{\perp}$. }}
\end{figure}

Because the slab targets are invariant in y near the central region
where the laser is incident, one might expect them to be essentially
well modeled by a 2D simulation and it is tempting to accept the physical
picture of electron trapping in the slabs presented above. But, as
we discussed earlier, the simulation problem is in fact, not 2D because
the laser fields provide forces on the electrons that have an essential
3D element, specifically forces in the y dimension. Fig. \textcolor{blue}{\ref{fig:track_GRT_SPK}
(a)} shows the trajectories of the high energy hot electrons from
the 3D simulation; in contrast to the 2D simulation, they do not propagate
along the structure surfaces.

\subsection{Electron Confinement in 3D}
\label{3D_confinement}

Fig. \textcolor{blue}{\ref{fig:physics_fin}(a)} indicates the responsible
physics: After the electrons are pulled out into the gap, in addition
to a $B_{y}$ field, there is also a $B_{x}$ field around the hot
electron current. This field pinches the hot electron current in the
y direction, leading to an increase in the current density $j_{hot}$
as it propagates. A plot of the B field in the simulation is shown
in Fig. \textcolor{blue}{\ref{fig:physics_fin}(b)}. It is an xy section
of $B_{\perp}$, the B field normal to z. The color map shows the
magnitude while the black arrows indicate the directions. While the
return current remains the same as in the 2D simulation, here the
$B_{y}$ field stemming from both $j_{hot}$ and $j_{return}$ grows
such that the Lorentz force exceeds the Coulomb force. Electrons are
thus pulled away from the slab surface, and since the spacing between
the slabs is much smaller than the length of the slabs, electrons
can easily hit the slab on the other side and again bounce back into
the gap. In general, the 3D simulation shows that the electrons are
pinched in the y dimension, and bounce back and forth in the x dimension.
For the electrons, each bounce off of a surface means a sudden reversal
of sign in $v_{x}$, and the DLA force, $-ev_{x}B_{laser}$, becomes
a decelerating force, effectively terminating the acceleration length
and thus restraining the energy gain in the forward direction. This
physics is responsible for the fact that the cutoff energy found in
our 3D simulations of the slabs is consistently smaller than the cutoff
energy found in our 2D simulations. In addition, these bounces also
widen the angular distribution of the emitted electrons in the x direction
while, in the y direction, the angular spread is determined by the
pinching. This explains the asymmetry in Fig. \textcolor{blue}{\ref{fig:ang_distrib_3D}(b2)}

\begin{figure}[pht]
\centering{}\includegraphics[width=80mm]{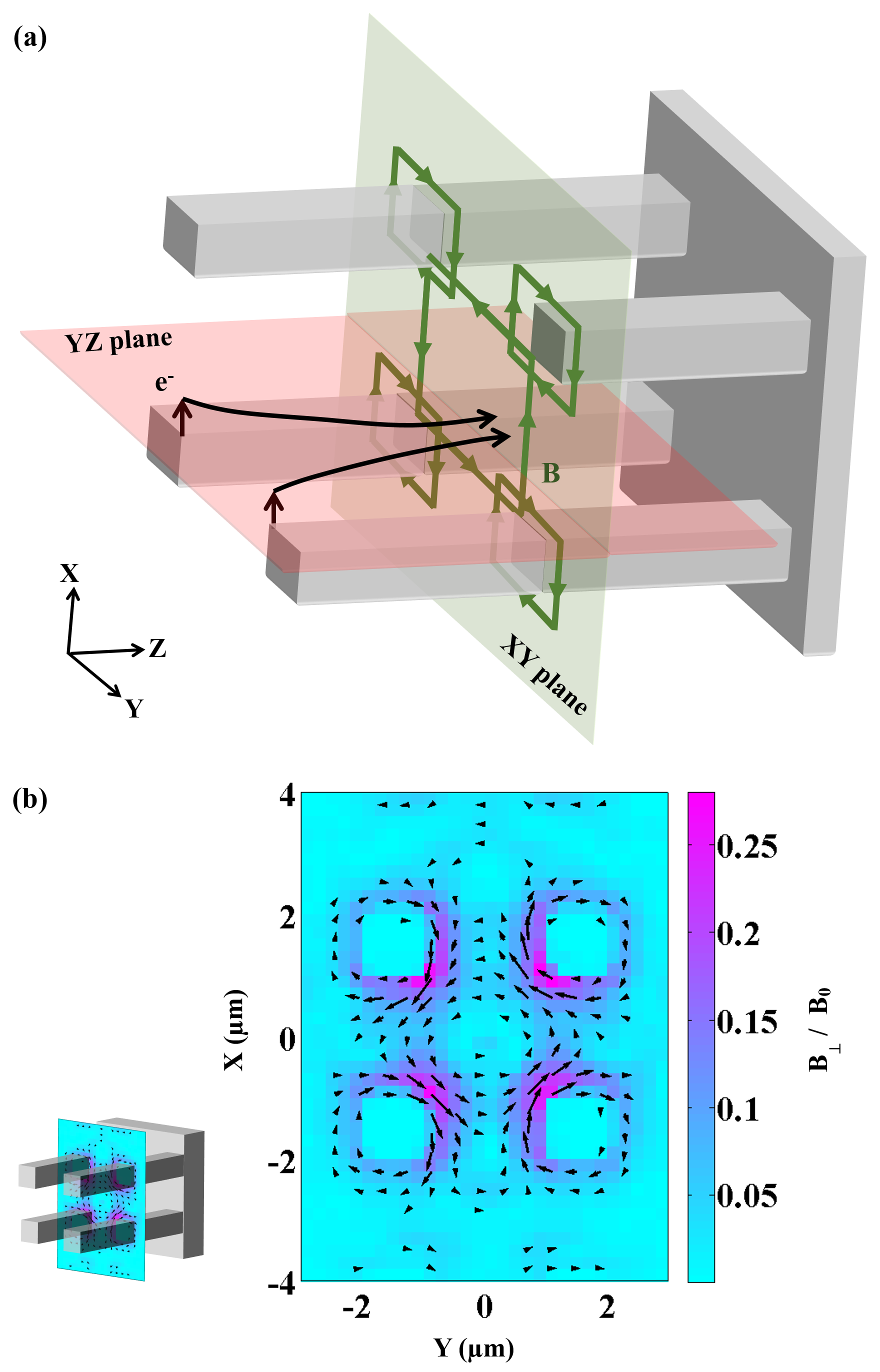} \caption{\label{fig:physics_spk} \emph{Guiding mechanisms in 3D towers. In
(}\textcolor{blue}{\emph{a}}\emph{), The quasi-static B field induced
by the hot electron current and the return current is indicated by
the green arrows. Since the laser polarization is in the x direction,
the electrons pulled out by the laser E field mainly fill in the gaps
in the XZ plane rather than the YZ plane. The $B_{x}$ field that
dominates in the gaps pulls the electrons along y towards the center.
In the x direction near the surfaces, the $E_{x}$ and $B_{y}$ fields
still balance out. So electrons are constrained in the x direction
and pinched in the y direction. Similar to Fig. }\textcolor{blue}{\emph{\ref{fig:physics_fin}(b)}}\emph{,
(}\textcolor{blue}{\emph{b}}\emph{) is a plot of $B_{\perp}$ at $Z=-5\mu m$
where the peak of the laser pulse is located.}}
\end{figure}

Because the 3D simulation of the slab targets suggests that the acceleration
length is constrained by the collisions with the walls, an obvious
improvement is to reduce the probability of collision with the walls
by replacing the slabs with towers (see Fig. \textcolor{blue}{\ref{fig:3struct_fE}}).
A field analysis of the tower target structure is shown in Fig. \textcolor{blue}{\ref{fig:physics_spk}}.
The green arrows in Fig. \textcolor{blue}{\ref{fig:physics_spk}(a)}
are a cartoon of the B field directions in the xy plane consistent
with the directions of $j_{hot}$ and $j_{return}$. Fig. \textcolor{blue}{\ref{fig:physics_spk}(b)}
is an xy field map taken from the simulation indicating $B_{\perp}$
at the plane where the laser peak is located. Electrons start off
by being pulled out by the laser E field in the x direction, where
they spill into the gaps in the XZ plane between the towers. Inside
these gaps, because the E field points mainly in the charge separation
direction x, its force on the electrons can balance that due to $cB_{y}$
when the electrons are along the tower surfaces. However, the $cB_{x}$
component of the Lorentz force causes the electrons to move in the
y direction towards y=0. The general trend of the high energy fast
electron trajectories is shown by the black curves in Fig. \textcolor{blue}{\ref{fig:physics_spk}(a)}.
Electrons are mostly confined in the yz plane (red plane in Fig.
\textcolor{blue}{\ref{fig:physics_spk}(a)}) close to the towers.
They are bent towards the center in the y direction, where the $B_{y}$
field is small and plays no important role in the electron trajectories.
This explains the two peaks along the y direction in the angular distribution
map Fig. \textcolor{blue}{\ref{fig:ang_distrib_3D}(c2)}.

Fig. \textcolor{blue}{\ref{fig:track_GRT_SPK}} shows the trajectories
of a sampling of electrons above $120MeV$ in simulations for both slab and tower
targets. Electrons originating from $X>0$ ($X<0$) are labeled red
(blue). For the slab-structured target, it is clear that electrons
bounce between the slab walls looking down the y axis (Fig. \textcolor{blue}{\ref{fig:track_GRT_SPK}(a2)}),
and pinch looking down the x axis (Fig. \textcolor{blue}{\ref{fig:track_GRT_SPK}(a3)}).
For the tower-structured target, the confinement is in the XZ plane
(Fig. \textcolor{blue}{\ref{fig:track_GRT_SPK}(b2)}) and the pinching is in the YZ plane (Fig. \textcolor{blue}{\ref{fig:track_GRT_SPK}(b3)}).

\section{Conclusion}

We have proposed front surface target structures (towers and slabs)
for use in the generation of high energy, collimated electrons. We
have shown that these targets cannot be adequately studied in 2D PIC
simulations, but require 3D. This will generally be the case for structured
targets. Compared to regular flat targets with $1\mu m$ pre-plasma,
the yield of electrons at the high energy end can be improved by several
orders of magnitude, while the FWHM divergence angle of the most energetic
hot electrons can be greatly reduced to $<$5$^{\circ}$. For electrons
above $100MeV$, the average current can be as high as $70kA$. For
high energy electrons, e.g. electrons at $100MeV$, the brightness
is on the order of $10^{23}s^{-1}mm^{-2}mrad^{-2}(0.1\%\ bandwidth)$.
The conversion efficiency from laser to fast electrons using towers
is comparable to that using a flat target with pre-plasma, while using
slabs the conversion efficiency is improved by $28\%$. For electrons
with energies above $50MeV$, the conversion efficiencies are $1.5\%$, $5.2\%$ and $7.9\%$ for the three targets respectively. This makes front surface
targets ideal for applications requiring collimated high-energy electrons.
We have shown that the hot electrons generated using our structured
targets are accelerated through the direct laser acceleration mechanism
(DLA) and are guided by surface fields. Since the guiding effect is
a sensitive function of the details of the target shape, a precise
description of the target shape in 3D is required not only to determine
the correct guiding fields, but is also important in terms of getting
correct background plasma density and wave guiding effects, both of
which can drastically change the electron energy. 
\begin{acknowledgments}
This work is supported by the AFOSR Young Investigator Program (YIP)
under contract No.FA9550-12-1-0341. Sheng Jiang would like to thank Dr. Morrison for useful discussions. Computational time was granted by the Ohio Supercomputer Center.
\end{acknowledgments}
\bibliography{bibfile_structure}

\begin{thebibliography}{48}
\expandafter\ifx\csname natexlab\endcsname\relax\def\natexlab#1{#1}\fi
\expandafter\ifx\csname bibnamefont\endcsname\relax
  \def\bibnamefont#1{#1}\fi
\expandafter\ifx\csname bibfnamefont\endcsname\relax
  \def\bibfnamefont#1{#1}\fi
\expandafter\ifx\csname citenamefont\endcsname\relax
  \def\citenamefont#1{#1}\fi
\expandafter\ifx\csname url\endcsname\relax
  \def\url#1{\texttt{#1}}\fi
\expandafter\ifx\csname urlprefix\endcsname\relax\def\urlprefix{URL }\fi
\providecommand{\bibinfo}[2]{#2}
\providecommand{\eprint}[2][]{\url{#2}}

\bibitem[{\citenamefont{Murnane et~al.}(1989)\citenamefont{Murnane, Kapteyn,
  and Falcone}}]{Murnane:1989}
\bibinfo{author}{\bibfnamefont{M.~M.} \bibnamefont{Murnane}},
  \bibinfo{author}{\bibfnamefont{H.~C.} \bibnamefont{Kapteyn}},
  \bibnamefont{and} \bibinfo{author}{\bibfnamefont{R.~W.}
  \bibnamefont{Falcone}}, \bibinfo{journal}{Phy. Rev. Lett.}
  \textbf{\bibinfo{volume}{62}}, \bibinfo{pages}{155} (\bibinfo{year}{1989}).

\bibitem[{\citenamefont{Kmetec et~al.}(1992)\citenamefont{Kmetec, Gordon,
  Macklin, Lemoff, Brown, and Harris}}]{Kmetec:1992}
\bibinfo{author}{\bibfnamefont{J.~D.} \bibnamefont{Kmetec}},
  \bibinfo{author}{\bibfnamefont{C.~L.} \bibnamefont{Gordon}},
  \bibinfo{author}{\bibfnamefont{J.~J.} \bibnamefont{Macklin}},
  \bibinfo{author}{\bibfnamefont{B.~E.} \bibnamefont{Lemoff}},
  \bibinfo{author}{\bibfnamefont{G.~S.} \bibnamefont{Brown}}, \bibnamefont{and}
  \bibinfo{author}{\bibfnamefont{S.~E.} \bibnamefont{Harris}},
  \bibinfo{journal}{Phy. Rev. Lett.} \textbf{\bibinfo{volume}{68}},
  \bibinfo{pages}{1527} (\bibinfo{year}{1992}).

\bibitem[{\citenamefont{Schnürer et~al.}(1995)\citenamefont{Schnürer,
  Kalashnikov, Nickles, Schlegel, Sandner, Demchenko, Nolte, and
  Ambrosi}}]{Schnurer:1995}
\bibinfo{author}{\bibfnamefont{M.}~\bibnamefont{Schnürer}},
  \bibinfo{author}{\bibfnamefont{M.~P.} \bibnamefont{Kalashnikov}},
  \bibinfo{author}{\bibfnamefont{P.~V.} \bibnamefont{Nickles}},
  \bibinfo{author}{\bibfnamefont{T.}~\bibnamefont{Schlegel}},
  \bibinfo{author}{\bibfnamefont{W.}~\bibnamefont{Sandner}},
  \bibinfo{author}{\bibfnamefont{N.}~\bibnamefont{Demchenko}},
  \bibinfo{author}{\bibfnamefont{R.}~\bibnamefont{Nolte}}, \bibnamefont{and}
  \bibinfo{author}{\bibfnamefont{P.}~\bibnamefont{Ambrosi}},
  \bibinfo{journal}{Phys. Plasmas} \textbf{\bibinfo{volume}{2}},
  \bibinfo{pages}{3106} (\bibinfo{year}{1995}).

\bibitem[{\citenamefont{Shearer et~al.}(1973)\citenamefont{Shearer, Garrison,
  Wong, and Swain}}]{Shearer:1973}
\bibinfo{author}{\bibfnamefont{J.~W.} \bibnamefont{Shearer}},
  \bibinfo{author}{\bibfnamefont{J.}~\bibnamefont{Garrison}},
  \bibinfo{author}{\bibfnamefont{J.}~\bibnamefont{Wong}}, \bibnamefont{and}
  \bibinfo{author}{\bibfnamefont{J.~E.} \bibnamefont{Swain}},
  \bibinfo{journal}{Phys. Rev. A} \textbf{\bibinfo{volume}{8}},
  \bibinfo{pages}{1582} (\bibinfo{year}{1973}).

\bibitem[{\citenamefont{Chen et~al.}(2009)\citenamefont{Chen, Wilks, Bonlie,
  Liang, Myatt, Price, Meyerhofer, and Beiersdorfer}}]{Chen:PRL2009}
\bibinfo{author}{\bibfnamefont{H.}~\bibnamefont{Chen}},
  \bibinfo{author}{\bibfnamefont{S.~C.} \bibnamefont{Wilks}},
  \bibinfo{author}{\bibfnamefont{J.~D.} \bibnamefont{Bonlie}},
  \bibinfo{author}{\bibfnamefont{E.~P.} \bibnamefont{Liang}},
  \bibinfo{author}{\bibfnamefont{J.}~\bibnamefont{Myatt}},
  \bibinfo{author}{\bibfnamefont{D.~F.} \bibnamefont{Price}},
  \bibinfo{author}{\bibfnamefont{D.~D.} \bibnamefont{Meyerhofer}},
  \bibnamefont{and}
  \bibinfo{author}{\bibfnamefont{P.}~\bibnamefont{Beiersdorfer}},
  \bibinfo{journal}{Phys. Rev. Lett.} \textbf{\bibinfo{volume}{102}},
  \bibinfo{pages}{105001} (\bibinfo{year}{2009}).

\bibitem[{\citenamefont{Clark et~al.}(2000)\citenamefont{Clark, Krushelnick,
  Davies, Zepf, Tatarakis, Beg, Machacek, Norreys, Santala, Watts
  et~al.}}]{Clark:2000}
\bibinfo{author}{\bibfnamefont{E.~L.} \bibnamefont{Clark}},
  \bibinfo{author}{\bibfnamefont{K.}~\bibnamefont{Krushelnick}},
  \bibinfo{author}{\bibfnamefont{J.~R.} \bibnamefont{Davies}},
  \bibinfo{author}{\bibfnamefont{M.}~\bibnamefont{Zepf}},
  \bibinfo{author}{\bibfnamefont{M.}~\bibnamefont{Tatarakis}},
  \bibinfo{author}{\bibfnamefont{F.~N.} \bibnamefont{Beg}},
  \bibinfo{author}{\bibfnamefont{A.}~\bibnamefont{Machacek}},
  \bibinfo{author}{\bibfnamefont{P.~A.} \bibnamefont{Norreys}},
  \bibinfo{author}{\bibfnamefont{M.~I.~K.} \bibnamefont{Santala}},
  \bibinfo{author}{\bibfnamefont{I.}~\bibnamefont{Watts}},
  \bibnamefont{et~al.}, \bibinfo{journal}{Phy. Rev. Lett.}
  \textbf{\bibinfo{volume}{84}}, \bibinfo{pages}{670} (\bibinfo{year}{2000}).

\bibitem[{\citenamefont{Wilks et~al.}(1992)\citenamefont{Wilks, Kruer, Tabak,
  and Langdon}}]{Wilks:1992}
\bibinfo{author}{\bibfnamefont{S.~C.} \bibnamefont{Wilks}},
  \bibinfo{author}{\bibfnamefont{W.~L.} \bibnamefont{Kruer}},
  \bibinfo{author}{\bibfnamefont{M.}~\bibnamefont{Tabak}}, \bibnamefont{and}
  \bibinfo{author}{\bibfnamefont{A.~B.} \bibnamefont{Langdon}},
  \bibinfo{journal}{Phy. Rev. Lett.} \textbf{\bibinfo{volume}{69}},
  \bibinfo{pages}{1383} (\bibinfo{year}{1992}).

\bibitem[{\citenamefont{Beg et~al.}(1997)\citenamefont{Beg, Bell, Dangor,
  Danson, Fews, Glinsky, Hammel, Lee, Norreys, and Tatarakis}}]{Beg:1997}
\bibinfo{author}{\bibfnamefont{F.~N.} \bibnamefont{Beg}},
  \bibinfo{author}{\bibfnamefont{A.~R.} \bibnamefont{Bell}},
  \bibinfo{author}{\bibfnamefont{A.~E.} \bibnamefont{Dangor}},
  \bibinfo{author}{\bibfnamefont{C.~N.} \bibnamefont{Danson}},
  \bibinfo{author}{\bibfnamefont{A.~P.} \bibnamefont{Fews}},
  \bibinfo{author}{\bibfnamefont{M.~E.} \bibnamefont{Glinsky}},
  \bibinfo{author}{\bibfnamefont{B.~A.} \bibnamefont{Hammel}},
  \bibinfo{author}{\bibfnamefont{P.}~\bibnamefont{Lee}},
  \bibinfo{author}{\bibfnamefont{P.~A.} \bibnamefont{Norreys}},
  \bibnamefont{and}
  \bibinfo{author}{\bibfnamefont{M.}~\bibnamefont{Tatarakis}},
  \bibinfo{journal}{Phys. Plasmas} \textbf{\bibinfo{volume}{4}},
  \bibinfo{pages}{447} (\bibinfo{year}{1997}).

\bibitem[{\citenamefont{Haines et~al.}(2009)\citenamefont{Haines, Wei, Beg, and
  Stephens}}]{Haines:2009}
\bibinfo{author}{\bibfnamefont{M.~G.} \bibnamefont{Haines}},
  \bibinfo{author}{\bibfnamefont{M.~S.} \bibnamefont{Wei}},
  \bibinfo{author}{\bibfnamefont{F.~N.} \bibnamefont{Beg}}, \bibnamefont{and}
  \bibinfo{author}{\bibfnamefont{R.~B.} \bibnamefont{Stephens}},
  \bibinfo{journal}{Phy. Rev. Lett.} \textbf{\bibinfo{volume}{102}},
  \bibinfo{pages}{045008} (\bibinfo{year}{2009}).

\bibitem[{\citenamefont{Kluge et~al.}(2011)\citenamefont{Kluge, Cowan, Debus,
  Schramm, Zeil, and Bussmann}}]{KlugeEnergy:2011}
\bibinfo{author}{\bibfnamefont{T.}~\bibnamefont{Kluge}},
  \bibinfo{author}{\bibfnamefont{T.}~\bibnamefont{Cowan}},
  \bibinfo{author}{\bibfnamefont{A.}~\bibnamefont{Debus}},
  \bibinfo{author}{\bibfnamefont{U.}~\bibnamefont{Schramm}},
  \bibinfo{author}{\bibfnamefont{K.}~\bibnamefont{Zeil}}, \bibnamefont{and}
  \bibinfo{author}{\bibfnamefont{M.}~\bibnamefont{Bussmann}},
  \bibinfo{journal}{Phy. Rev. Lett.} \textbf{\bibinfo{volume}{107}},
  \bibinfo{pages}{205003} (\bibinfo{year}{2011}).

\bibitem[{\citenamefont{Paradkar et~al.}(2011)\citenamefont{Paradkar, Wei,
  Yabuuchi, Stephens, Haines, Krasheninnikov, and Beg}}]{Paradkar:2011}
\bibinfo{author}{\bibfnamefont{B.~S.} \bibnamefont{Paradkar}},
  \bibinfo{author}{\bibfnamefont{M.~S.} \bibnamefont{Wei}},
  \bibinfo{author}{\bibfnamefont{T.}~\bibnamefont{Yabuuchi}},
  \bibinfo{author}{\bibfnamefont{R.~B.} \bibnamefont{Stephens}},
  \bibinfo{author}{\bibfnamefont{M.~G.} \bibnamefont{Haines}},
  \bibinfo{author}{\bibfnamefont{S.~I.} \bibnamefont{Krasheninnikov}},
  \bibnamefont{and} \bibinfo{author}{\bibfnamefont{F.~N.} \bibnamefont{Beg}},
  \bibinfo{journal}{Phys. Rev. E} \textbf{\bibinfo{volume}{83}},
  \bibinfo{pages}{046401} (\bibinfo{year}{2011}).

\bibitem[{\citenamefont{Scott et~al.}(2012)\citenamefont{Scott, Perez, Santos,
  and Ridgers}}]{Scott:2012}
\bibinfo{author}{\bibfnamefont{R.~H.~H.} \bibnamefont{Scott}},
  \bibinfo{author}{\bibfnamefont{F.}~\bibnamefont{Perez}},
  \bibinfo{author}{\bibfnamefont{J.~J.} \bibnamefont{Santos}},
  \bibnamefont{and} \bibinfo{author}{\bibfnamefont{C.~P.}
  \bibnamefont{Ridgers}}, \bibinfo{journal}{Phys. Plasmas}
  \textbf{\bibinfo{volume}{19}}, \bibinfo{pages}{053104}
  (\bibinfo{year}{2012}).

\bibitem[{\citenamefont{Ovchinnikov et~al.}(2013)\citenamefont{Ovchinnikov,
  Schumacher, McMahon, Chowdhury, Chen, Morace, and
  Freeman}}]{Ovchinnikov:2013}
\bibinfo{author}{\bibfnamefont{V.~M.} \bibnamefont{Ovchinnikov}},
  \bibinfo{author}{\bibfnamefont{D.~W.} \bibnamefont{Schumacher}},
  \bibinfo{author}{\bibfnamefont{M.}~\bibnamefont{McMahon}},
  \bibinfo{author}{\bibfnamefont{E.~A.} \bibnamefont{Chowdhury}},
  \bibinfo{author}{\bibfnamefont{C.~D.} \bibnamefont{Chen}},
  \bibinfo{author}{\bibfnamefont{A.}~\bibnamefont{Morace}}, \bibnamefont{and}
  \bibinfo{author}{\bibfnamefont{R.~R.} \bibnamefont{Freeman}},
  \bibinfo{journal}{Phy. Rev. Lett.} \textbf{\bibinfo{volume}{110}},
  \bibinfo{pages}{065007} (\bibinfo{year}{2013}).

\bibitem[{\citenamefont{Palchan et~al.}(2007)\citenamefont{Palchan, Pecker,
  Henis, Eisenmann, and Zigler}}]{Palchan:2007}
\bibinfo{author}{\bibfnamefont{T.}~\bibnamefont{Palchan}},
  \bibinfo{author}{\bibfnamefont{S.}~\bibnamefont{Pecker}},
  \bibinfo{author}{\bibfnamefont{Z.}~\bibnamefont{Henis}},
  \bibinfo{author}{\bibfnamefont{S.}~\bibnamefont{Eisenmann}},
  \bibnamefont{and} \bibinfo{author}{\bibfnamefont{A.}~\bibnamefont{Zigler}},
  \bibinfo{journal}{App. Phys. Lett.} \textbf{\bibinfo{volume}{90}},
  \bibinfo{pages}{041501} (\bibinfo{year}{2007}).

\bibitem[{\citenamefont{Rajeev et~al.}(2003)\citenamefont{Rajeev, Taneja,
  Ayyub, Sandhu, and Kumar}}]{Rajeev:2003}
\bibinfo{author}{\bibfnamefont{P.~P.} \bibnamefont{Rajeev}},
  \bibinfo{author}{\bibfnamefont{P.}~\bibnamefont{Taneja}},
  \bibinfo{author}{\bibfnamefont{P.}~\bibnamefont{Ayyub}},
  \bibinfo{author}{\bibfnamefont{A.~S.} \bibnamefont{Sandhu}},
  \bibnamefont{and} \bibinfo{author}{\bibfnamefont{G.~R.} \bibnamefont{Kumar}},
  \bibinfo{journal}{Phy. Rev. Lett.} \textbf{\bibinfo{volume}{90}},
  \bibinfo{pages}{115002} (\bibinfo{year}{2003}).

\bibitem[{\citenamefont{Sumeruk et~al.}(2007)\citenamefont{Sumeruk, Kneip,
  Symes, Churina, Belolipetski, Donnelly, and Ditmire}}]{Sumeruk:2007}
\bibinfo{author}{\bibfnamefont{H.~A.} \bibnamefont{Sumeruk}},
  \bibinfo{author}{\bibfnamefont{S.}~\bibnamefont{Kneip}},
  \bibinfo{author}{\bibfnamefont{D.~R.} \bibnamefont{Symes}},
  \bibinfo{author}{\bibfnamefont{I.~V.} \bibnamefont{Churina}},
  \bibinfo{author}{\bibfnamefont{A.~V.} \bibnamefont{Belolipetski}},
  \bibinfo{author}{\bibfnamefont{T.~D.} \bibnamefont{Donnelly}},
  \bibnamefont{and} \bibinfo{author}{\bibfnamefont{T.}~\bibnamefont{Ditmire}},
  \bibinfo{journal}{Phy. Rev. Lett.} \textbf{\bibinfo{volume}{98}},
  \bibinfo{pages}{045001} (\bibinfo{year}{2007}).

\bibitem[{\citenamefont{Kulcsar et~al.}(2000)\citenamefont{Kulcsar, AlMawlawi,
  Budnik, Herman, Moskovits, Zhao, and Marjoribanks}}]{Kulcsar:2000}
\bibinfo{author}{\bibfnamefont{G.}~\bibnamefont{Kulcsar}},
  \bibinfo{author}{\bibfnamefont{D.}~\bibnamefont{AlMawlawi}},
  \bibinfo{author}{\bibfnamefont{F.~W.} \bibnamefont{Budnik}},
  \bibinfo{author}{\bibfnamefont{P.~R.} \bibnamefont{Herman}},
  \bibinfo{author}{\bibfnamefont{M.}~\bibnamefont{Moskovits}},
  \bibinfo{author}{\bibfnamefont{L.}~\bibnamefont{Zhao}}, \bibnamefont{and}
  \bibinfo{author}{\bibfnamefont{R.~S.} \bibnamefont{Marjoribanks}},
  \bibinfo{journal}{Phy. Rev. Lett.} \textbf{\bibinfo{volume}{84}},
  \bibinfo{pages}{5149} (\bibinfo{year}{2000}).

\bibitem[{\citenamefont{Murnane et~al.}(1993)\citenamefont{Murnane, Kapteyn,
  Gordon, Bokor, and Glytsis}}]{Murnane:1993}
\bibinfo{author}{\bibfnamefont{M.~M.} \bibnamefont{Murnane}},
  \bibinfo{author}{\bibfnamefont{H.~C.} \bibnamefont{Kapteyn}},
  \bibinfo{author}{\bibfnamefont{S.~P.} \bibnamefont{Gordon}},
  \bibinfo{author}{\bibfnamefont{J.}~\bibnamefont{Bokor}}, \bibnamefont{and}
  \bibinfo{author}{\bibfnamefont{E.~N.} \bibnamefont{Glytsis}},
  \bibinfo{journal}{App. Phys. Lett.} \textbf{\bibinfo{volume}{62}},
  \bibinfo{pages}{1068} (\bibinfo{year}{1993}).

\bibitem[{\citenamefont{Kahaly et~al.}(2008)\citenamefont{Kahaly, Yadav, Wang,
  Sengupta, Sheng, Das, Kaw, and Kumar}}]{Kahaly:PRL2008}
\bibinfo{author}{\bibfnamefont{S.}~\bibnamefont{Kahaly}},
  \bibinfo{author}{\bibfnamefont{S.~K.} \bibnamefont{Yadav}},
  \bibinfo{author}{\bibfnamefont{W.~M.} \bibnamefont{Wang}},
  \bibinfo{author}{\bibfnamefont{S.}~\bibnamefont{Sengupta}},
  \bibinfo{author}{\bibfnamefont{Z.~M.} \bibnamefont{Sheng}},
  \bibinfo{author}{\bibfnamefont{A.}~\bibnamefont{Das}},
  \bibinfo{author}{\bibfnamefont{P.~K.} \bibnamefont{Kaw}}, \bibnamefont{and}
  \bibinfo{author}{\bibfnamefont{G.~R.} \bibnamefont{Kumar}},
  \bibinfo{journal}{Phy. Rev. Lett.} \textbf{\bibinfo{volume}{101}},
  \bibinfo{pages}{145001} (\bibinfo{year}{2008}).

\bibitem[{\citenamefont{Hu et~al.}(2010)\citenamefont{Hu, Lei, Wang, Huang,
  Wang, Wang, Xu, Shen, Liu, Yu et~al.}}]{Hu:POP2010}
\bibinfo{author}{\bibfnamefont{G.}~\bibnamefont{Hu}},
  \bibinfo{author}{\bibfnamefont{A.}~\bibnamefont{Lei}},
  \bibinfo{author}{\bibfnamefont{J.}~\bibnamefont{Wang}},
  \bibinfo{author}{\bibfnamefont{L.}~\bibnamefont{Huang}},
  \bibinfo{author}{\bibfnamefont{W.}~\bibnamefont{Wang}},
  \bibinfo{author}{\bibfnamefont{X.}~\bibnamefont{Wang}},
  \bibinfo{author}{\bibfnamefont{Y.}~\bibnamefont{Xu}},
  \bibinfo{author}{\bibfnamefont{B.}~\bibnamefont{Shen}},
  \bibinfo{author}{\bibfnamefont{J.}~\bibnamefont{Liu}},
  \bibinfo{author}{\bibfnamefont{W.}~\bibnamefont{Yu}}, \bibnamefont{et~al.},
  \bibinfo{journal}{Phys. Plasmas} \textbf{\bibinfo{volume}{17}},
  \bibinfo{pages}{083102} (\bibinfo{year}{2010}).

\bibitem[{\citenamefont{Margarone et~al.}(2012)\citenamefont{Margarone, Klimo,
  Kim, Prokupek, Limpouch, Jeong, Mocek, Pšikal, Kim, Proška
  et~al.}}]{Margarone:PRL2012}
\bibinfo{author}{\bibfnamefont{D.}~\bibnamefont{Margarone}},
  \bibinfo{author}{\bibfnamefont{O.}~\bibnamefont{Klimo}},
  \bibinfo{author}{\bibfnamefont{I.~J.} \bibnamefont{Kim}},
  \bibinfo{author}{\bibfnamefont{J.}~\bibnamefont{Prokupek}},
  \bibinfo{author}{\bibfnamefont{J.}~\bibnamefont{Limpouch}},
  \bibinfo{author}{\bibfnamefont{T.~M.} \bibnamefont{Jeong}},
  \bibinfo{author}{\bibfnamefont{T.}~\bibnamefont{Mocek}},
  \bibinfo{author}{\bibfnamefont{J.}~\bibnamefont{Pšikal}},
  \bibinfo{author}{\bibfnamefont{H.~T.} \bibnamefont{Kim}},
  \bibinfo{author}{\bibfnamefont{J.}~\bibnamefont{Proška}},
  \bibnamefont{et~al.}, \bibinfo{journal}{Phy. Rev. Lett.}
  \textbf{\bibinfo{volume}{109}}, \bibinfo{pages}{234801}
  (\bibinfo{year}{2012}).

\bibitem[{\citenamefont{Kupersztych et~al.}(2004)\citenamefont{Kupersztych,
  Raynaud, and Riconda}}]{Kupersztych:POP2004}
\bibinfo{author}{\bibfnamefont{J.}~\bibnamefont{Kupersztych}},
  \bibinfo{author}{\bibfnamefont{M.}~\bibnamefont{Raynaud}}, \bibnamefont{and}
  \bibinfo{author}{\bibfnamefont{C.}~\bibnamefont{Riconda}},
  \bibinfo{journal}{Phys. Plasmas} \textbf{\bibinfo{volume}{11}},
  \bibinfo{pages}{1669} (\bibinfo{year}{2004}).

\bibitem[{\citenamefont{Klimo et~al.}(2011)\citenamefont{Klimo, Psikal,
  Limpouch, Proska, Novotny, Ceccotti, Floquet, and Kawata}}]{Klimo:NJP2011}
\bibinfo{author}{\bibfnamefont{O.}~\bibnamefont{Klimo}},
  \bibinfo{author}{\bibfnamefont{J.}~\bibnamefont{Psikal}},
  \bibinfo{author}{\bibfnamefont{J.}~\bibnamefont{Limpouch}},
  \bibinfo{author}{\bibfnamefont{J.}~\bibnamefont{Proska}},
  \bibinfo{author}{\bibfnamefont{F.}~\bibnamefont{Novotny}},
  \bibinfo{author}{\bibfnamefont{T.}~\bibnamefont{Ceccotti}},
  \bibinfo{author}{\bibfnamefont{V.}~\bibnamefont{Floquet}}, \bibnamefont{and}
  \bibinfo{author}{\bibfnamefont{S.}~\bibnamefont{Kawata}},
  \bibinfo{journal}{New J. Phys.} \textbf{\bibinfo{volume}{13}},
  \bibinfo{pages}{053028} (\bibinfo{year}{2011}).

\bibitem[{\citenamefont{Andreev et~al.}(2011)\citenamefont{Andreev, Kumar,
  Platonov, and Pukhov}}]{Andreev:POP2011}
\bibinfo{author}{\bibfnamefont{A.}~\bibnamefont{Andreev}},
  \bibinfo{author}{\bibfnamefont{N.}~\bibnamefont{Kumar}},
  \bibinfo{author}{\bibfnamefont{K.}~\bibnamefont{Platonov}}, \bibnamefont{and}
  \bibinfo{author}{\bibfnamefont{A.}~\bibnamefont{Pukhov}},
  \bibinfo{journal}{Phys. Plasmas} \textbf{\bibinfo{volume}{18}},
  \bibinfo{pages}{103103} (\bibinfo{year}{2011}).

\bibitem[{\citenamefont{Kemp et~al.}(2013)\citenamefont{Kemp, Link, Ping,
  Schumacher, Freeman, and Patel}}]{Kemp:POP2013}
\bibinfo{author}{\bibfnamefont{G.~E.} \bibnamefont{Kemp}},
  \bibinfo{author}{\bibfnamefont{A.}~\bibnamefont{Link}},
  \bibinfo{author}{\bibfnamefont{Y.}~\bibnamefont{Ping}},
  \bibinfo{author}{\bibfnamefont{D.~W.} \bibnamefont{Schumacher}},
  \bibinfo{author}{\bibfnamefont{R.~R.} \bibnamefont{Freeman}},
  \bibnamefont{and} \bibinfo{author}{\bibfnamefont{P.~K.} \bibnamefont{Patel}},
  \bibinfo{journal}{Phys. Plasmas} \textbf{\bibinfo{volume}{20}},
  \bibinfo{pages}{2033104} (\bibinfo{year}{2013}).

\bibitem[{\citenamefont{Kluge et~al.}(2012)\citenamefont{Kluge, Gaillard,
  Flippo, Burris-Mog, Enghardt, Gall, Geissel, Helm, Kraft, Lockard
  et~al.}}]{Kluge:2012}
\bibinfo{author}{\bibfnamefont{T.}~\bibnamefont{Kluge}},
  \bibinfo{author}{\bibfnamefont{S.~A.} \bibnamefont{Gaillard}},
  \bibinfo{author}{\bibfnamefont{K.~A.} \bibnamefont{Flippo}},
  \bibinfo{author}{\bibfnamefont{T.}~\bibnamefont{Burris-Mog}},
  \bibinfo{author}{\bibfnamefont{W.}~\bibnamefont{Enghardt}},
  \bibinfo{author}{\bibfnamefont{B.}~\bibnamefont{Gall}},
  \bibinfo{author}{\bibfnamefont{M.}~\bibnamefont{Geissel}},
  \bibinfo{author}{\bibfnamefont{A.}~\bibnamefont{Helm}},
  \bibinfo{author}{\bibfnamefont{S.~D.} \bibnamefont{Kraft}},
  \bibinfo{author}{\bibfnamefont{T.}~\bibnamefont{Lockard}},
  \bibnamefont{et~al.}, \bibinfo{journal}{New J. Phys.}
  \textbf{\bibinfo{volume}{14}}, \bibinfo{pages}{023038}
  (\bibinfo{year}{2012}).

\bibitem[{\citenamefont{Gaillard et~al.}(2011)\citenamefont{Gaillard, Kluge,
  Flippo, Bussmann, Gall, Lockard, Geissel, Offermann, Schollmeier, Sentoku
  et~al.}}]{Gaillard:POP2011}
\bibinfo{author}{\bibfnamefont{S.~A.} \bibnamefont{Gaillard}},
  \bibinfo{author}{\bibfnamefont{T.}~\bibnamefont{Kluge}},
  \bibinfo{author}{\bibfnamefont{K.~A.} \bibnamefont{Flippo}},
  \bibinfo{author}{\bibfnamefont{M.}~\bibnamefont{Bussmann}},
  \bibinfo{author}{\bibfnamefont{B.}~\bibnamefont{Gall}},
  \bibinfo{author}{\bibfnamefont{T.}~\bibnamefont{Lockard}},
  \bibinfo{author}{\bibfnamefont{M.}~\bibnamefont{Geissel}},
  \bibinfo{author}{\bibfnamefont{D.~T.} \bibnamefont{Offermann}},
  \bibinfo{author}{\bibfnamefont{M.}~\bibnamefont{Schollmeier}},
  \bibinfo{author}{\bibfnamefont{Y.}~\bibnamefont{Sentoku}},
  \bibnamefont{et~al.}, \bibinfo{journal}{Phys. Plasmas}
  \textbf{\bibinfo{volume}{18}}, \bibinfo{pages}{056710}
  (\bibinfo{year}{2011}).

\bibitem[{\citenamefont{Zigler et~al.}(2013)\citenamefont{Zigler, Eisenman,
  Botton, Nahum, Schleifer, Baspaly, Pomerantz, Abicht, Branzel, Priebe
  et~al.}}]{Zigler:2013}
\bibinfo{author}{\bibfnamefont{A.}~\bibnamefont{Zigler}},
  \bibinfo{author}{\bibfnamefont{S.}~\bibnamefont{Eisenman}},
  \bibinfo{author}{\bibfnamefont{M.}~\bibnamefont{Botton}},
  \bibinfo{author}{\bibfnamefont{E.}~\bibnamefont{Nahum}},
  \bibinfo{author}{\bibfnamefont{E.}~\bibnamefont{Schleifer}},
  \bibinfo{author}{\bibfnamefont{A.}~\bibnamefont{Baspaly}},
  \bibinfo{author}{\bibfnamefont{I.}~\bibnamefont{Pomerantz}},
  \bibinfo{author}{\bibfnamefont{F.}~\bibnamefont{Abicht}},
  \bibinfo{author}{\bibfnamefont{J.}~\bibnamefont{Branzel}},
  \bibinfo{author}{\bibfnamefont{G.}~\bibnamefont{Priebe}},
  \bibnamefont{et~al.}, \bibinfo{journal}{Phy. Rev. Lett.}
  \textbf{\bibinfo{volume}{110}}, \bibinfo{pages}{215004}
  (\bibinfo{year}{2013}).

\bibitem[{\citenamefont{Zheng et~al.}(2011)\citenamefont{Zheng, Sheng, Liu,
  Zhou, Xu, and Zhang}}]{Zheng:POP2011}
\bibinfo{author}{\bibfnamefont{J.}~\bibnamefont{Zheng}},
  \bibinfo{author}{\bibfnamefont{Z.-M.} \bibnamefont{Sheng}},
  \bibinfo{author}{\bibfnamefont{J.-L.} \bibnamefont{Liu}},
  \bibinfo{author}{\bibfnamefont{W.-M.} \bibnamefont{Zhou}},
  \bibinfo{author}{\bibfnamefont{H.}~\bibnamefont{Xu}}, \bibnamefont{and}
  \bibinfo{author}{\bibfnamefont{J.}~\bibnamefont{Zhang}},
  \bibinfo{journal}{Phys. Plasmas} \textbf{\bibinfo{volume}{18}},
  \bibinfo{pages}{113103} (\bibinfo{year}{2011}).

\bibitem[{\citenamefont{Cao et~al.}(2010{\natexlab{a}})\citenamefont{Cao, Gu,
  Zhao, Cao, Huang, Zhou, He, Yu, and Yu}}]{Cao:POP2010_1}
\bibinfo{author}{\bibfnamefont{L.}~\bibnamefont{Cao}},
  \bibinfo{author}{\bibfnamefont{Y.}~\bibnamefont{Gu}},
  \bibinfo{author}{\bibfnamefont{Z.}~\bibnamefont{Zhao}},
  \bibinfo{author}{\bibfnamefont{L.}~\bibnamefont{Cao}},
  \bibinfo{author}{\bibfnamefont{W.}~\bibnamefont{Huang}},
  \bibinfo{author}{\bibfnamefont{W.}~\bibnamefont{Zhou}},
  \bibinfo{author}{\bibfnamefont{X.~T.} \bibnamefont{He}},
  \bibinfo{author}{\bibfnamefont{W.}~\bibnamefont{Yu}}, \bibnamefont{and}
  \bibinfo{author}{\bibfnamefont{M.~Y.} \bibnamefont{Yu}},
  \bibinfo{journal}{Phys. Plasmas} \textbf{\bibinfo{volume}{17}},
  \bibinfo{pages}{043103} (\bibinfo{year}{2010}{\natexlab{a}}).

\bibitem[{\citenamefont{Cao et~al.}(2010{\natexlab{b}})\citenamefont{Cao, Gu,
  Zhao, Cao, Huang, Zhou, Cai, He, Yu, and Yu}}]{Cao:POP2010_2}
\bibinfo{author}{\bibfnamefont{L.}~\bibnamefont{Cao}},
  \bibinfo{author}{\bibfnamefont{Y.}~\bibnamefont{Gu}},
  \bibinfo{author}{\bibfnamefont{Z.}~\bibnamefont{Zhao}},
  \bibinfo{author}{\bibfnamefont{L.}~\bibnamefont{Cao}},
  \bibinfo{author}{\bibfnamefont{W.}~\bibnamefont{Huang}},
  \bibinfo{author}{\bibfnamefont{W.}~\bibnamefont{Zhou}},
  \bibinfo{author}{\bibfnamefont{H.~B.} \bibnamefont{Cai}},
  \bibinfo{author}{\bibfnamefont{X.~T.} \bibnamefont{He}},
  \bibinfo{author}{\bibfnamefont{W.}~\bibnamefont{Yu}}, \bibnamefont{and}
  \bibinfo{author}{\bibfnamefont{M.~Y.} \bibnamefont{Yu}},
  \bibinfo{journal}{Phys. Plasmas} \textbf{\bibinfo{volume}{17}},
  \bibinfo{pages}{103106} (\bibinfo{year}{2010}{\natexlab{b}}).

\bibitem[{\citenamefont{Zhao et~al.}(2010)\citenamefont{Zhao, Cao, Cao, Wang,
  Huang, Jiang, He, Wu, Zhu, Dong et~al.}}]{Zhao:POP2010}
\bibinfo{author}{\bibfnamefont{Z.}~\bibnamefont{Zhao}},
  \bibinfo{author}{\bibfnamefont{L.}~\bibnamefont{Cao}},
  \bibinfo{author}{\bibfnamefont{L.}~\bibnamefont{Cao}},
  \bibinfo{author}{\bibfnamefont{J.}~\bibnamefont{Wang}},
  \bibinfo{author}{\bibfnamefont{W.}~\bibnamefont{Huang}},
  \bibinfo{author}{\bibfnamefont{W.}~\bibnamefont{Jiang}},
  \bibinfo{author}{\bibfnamefont{Y.}~\bibnamefont{He}},
  \bibinfo{author}{\bibfnamefont{Y.}~\bibnamefont{Wu}},
  \bibinfo{author}{\bibfnamefont{B.}~\bibnamefont{Zhu}},
  \bibinfo{author}{\bibfnamefont{K.}~\bibnamefont{Dong}}, \bibnamefont{et~al.},
  \bibinfo{journal}{Phys. Plasmas} \textbf{\bibinfo{volume}{17}},
  \bibinfo{pages}{123108} (\bibinfo{year}{2010}).

\bibitem[{\citenamefont{Cao et~al.}(2011)\citenamefont{Cao, Chen, Zhao, Cai, Wu
  et~al.}}]{Cao:POP2011}
\bibinfo{author}{\bibfnamefont{L.}~\bibnamefont{Cao}},
  \bibinfo{author}{\bibfnamefont{M.}~\bibnamefont{Chen}},
  \bibinfo{author}{\bibfnamefont{Z.}~\bibnamefont{Zhao}},
  \bibinfo{author}{\bibfnamefont{H.}~\bibnamefont{Cai}},
  \bibinfo{author}{\bibfnamefont{S.}~\bibnamefont{Wu}}, \bibnamefont{et~al.},
  \bibinfo{journal}{Phys. Plasmas} \textbf{\bibinfo{volume}{18}},
  \bibinfo{pages}{054501} (\bibinfo{year}{2011}).

\bibitem[{\citenamefont{Wang et~al.}(2012)\citenamefont{Wang, Cao, Zhao, Yu,
  Gu, and He}}]{Wang:2012}
\bibinfo{author}{\bibfnamefont{H.}~\bibnamefont{Wang}},
  \bibinfo{author}{\bibfnamefont{L.}~\bibnamefont{Cao}},
  \bibinfo{author}{\bibfnamefont{Z.}~\bibnamefont{Zhao}},
  \bibinfo{author}{\bibfnamefont{M.~Y.} \bibnamefont{Yu}},
  \bibinfo{author}{\bibfnamefont{Y.}~\bibnamefont{Gu}}, \bibnamefont{and}
  \bibinfo{author}{\bibfnamefont{X.~T.} \bibnamefont{He}},
  \bibinfo{journal}{Laser and Particle Beams} \textbf{\bibinfo{volume}{30}},
  \bibinfo{pages}{553} (\bibinfo{year}{2012}).

\bibitem[{\citenamefont{Yu et~al.}(2012{\natexlab{a}})\citenamefont{Yu, Zhao,
  Jin, Wu, Yan et~al.}}]{Yu:POP2012}
\bibinfo{author}{\bibfnamefont{J.}~\bibnamefont{Yu}},
  \bibinfo{author}{\bibfnamefont{Z.}~\bibnamefont{Zhao}},
  \bibinfo{author}{\bibfnamefont{X.}~\bibnamefont{Jin}},
  \bibinfo{author}{\bibfnamefont{F.}~\bibnamefont{Wu}},
  \bibinfo{author}{\bibfnamefont{Y.}~\bibnamefont{Yan}}, \bibnamefont{et~al.},
  \bibinfo{journal}{Phys. Plasmas} \textbf{\bibinfo{volume}{19}},
  \bibinfo{pages}{053108} (\bibinfo{year}{2012}{\natexlab{a}}).

\bibitem[{\citenamefont{Yu et~al.}(2012{\natexlab{b}})\citenamefont{Yu, Zhou,
  Cao, Zhao, Cao, Li, and Gu}}]{Yu:APL2012}
\bibinfo{author}{\bibfnamefont{J.}~\bibnamefont{Yu}},
  \bibinfo{author}{\bibfnamefont{W.}~\bibnamefont{Zhou}},
  \bibinfo{author}{\bibfnamefont{L.}~\bibnamefont{Cao}},
  \bibinfo{author}{\bibfnamefont{Z.}~\bibnamefont{Zhao}},
  \bibinfo{author}{\bibfnamefont{L.}~\bibnamefont{Cao}},
  \bibinfo{author}{\bibfnamefont{B.}~\bibnamefont{Li}}, \bibnamefont{and}
  \bibinfo{author}{\bibfnamefont{Y.}~\bibnamefont{Gu}}, \bibinfo{journal}{App.
  Phys. Lett.} \textbf{\bibinfo{volume}{100}}, \bibinfo{pages}{204101}
  (\bibinfo{year}{2012}{\natexlab{b}}).

\bibitem[{\citenamefont{Baton et~al.}(2008)\citenamefont{Baton, Koenig, Fuchs,
  Benuzzi-Mounaix, Guillou, Loupias, Vinci, Gremillet, Rousseaux, Drouin
  et~al.}}]{Baton:POP2008}
\bibinfo{author}{\bibfnamefont{S.~D.} \bibnamefont{Baton}},
  \bibinfo{author}{\bibfnamefont{M.}~\bibnamefont{Koenig}},
  \bibinfo{author}{\bibfnamefont{J.}~\bibnamefont{Fuchs}},
  \bibinfo{author}{\bibfnamefont{A.}~\bibnamefont{Benuzzi-Mounaix}},
  \bibinfo{author}{\bibfnamefont{P.}~\bibnamefont{Guillou}},
  \bibinfo{author}{\bibfnamefont{B.}~\bibnamefont{Loupias}},
  \bibinfo{author}{\bibfnamefont{T.}~\bibnamefont{Vinci}},
  \bibinfo{author}{\bibfnamefont{L.}~\bibnamefont{Gremillet}},
  \bibinfo{author}{\bibfnamefont{C.}~\bibnamefont{Rousseaux}},
  \bibinfo{author}{\bibfnamefont{M.}~\bibnamefont{Drouin}},
  \bibnamefont{et~al.}, \bibinfo{journal}{Phys. Plasmas}
  \textbf{\bibinfo{volume}{15}}, \bibinfo{pages}{042706}
  (\bibinfo{year}{2008}).

\bibitem[{\citenamefont{Psikal et~al.}(2010)\citenamefont{Psikal, Tikhonchuk,
  Limpouch, and Klimo}}]{Psikal:POP2010}
\bibinfo{author}{\bibfnamefont{J.}~\bibnamefont{Psikal}},
  \bibinfo{author}{\bibfnamefont{V.~T.} \bibnamefont{Tikhonchuk}},
  \bibinfo{author}{\bibfnamefont{J.}~\bibnamefont{Limpouch}}, \bibnamefont{and}
  \bibinfo{author}{\bibfnamefont{O.}~\bibnamefont{Klimo}},
  \bibinfo{journal}{Phys. Plasmas} \textbf{\bibinfo{volume}{17}},
  \bibinfo{pages}{013102} (\bibinfo{year}{2010}).

\bibitem[{\citenamefont{Micheau et~al.}(2010)\citenamefont{Micheau, Debayle,
  d’Humières, Honrubia, Qiao, Zepf, Borghesi, and
  Geissler}}]{Micheau:POP2010}
\bibinfo{author}{\bibfnamefont{S.}~\bibnamefont{Micheau}},
  \bibinfo{author}{\bibfnamefont{A.}~\bibnamefont{Debayle}},
  \bibinfo{author}{\bibfnamefont{E.}~\bibnamefont{d’Humières}},
  \bibinfo{author}{\bibfnamefont{J.~J.} \bibnamefont{Honrubia}},
  \bibinfo{author}{\bibfnamefont{B.}~\bibnamefont{Qiao}},
  \bibinfo{author}{\bibfnamefont{M.}~\bibnamefont{Zepf}},
  \bibinfo{author}{\bibfnamefont{M.}~\bibnamefont{Borghesi}}, \bibnamefont{and}
  \bibinfo{author}{\bibfnamefont{M.}~\bibnamefont{Geissler}},
  \bibinfo{journal}{Phys. Plasmas} \textbf{\bibinfo{volume}{17}},
  \bibinfo{pages}{122703} (\bibinfo{year}{2010}).

\bibitem[{\citenamefont{Stephens et~al.}(2004)\citenamefont{Stephens, Snavely,
  Aglitskiy, Amiranoff, Andersen, Batani, Baton, Cowan, Freeman, Hall
  et~al.}}]{Stephens:PRE2004}
\bibinfo{author}{\bibfnamefont{R.~B.} \bibnamefont{Stephens}},
  \bibinfo{author}{\bibfnamefont{R.~A.} \bibnamefont{Snavely}},
  \bibinfo{author}{\bibfnamefont{Y.}~\bibnamefont{Aglitskiy}},
  \bibinfo{author}{\bibfnamefont{F.}~\bibnamefont{Amiranoff}},
  \bibinfo{author}{\bibfnamefont{C.}~\bibnamefont{Andersen}},
  \bibinfo{author}{\bibfnamefont{D.}~\bibnamefont{Batani}},
  \bibinfo{author}{\bibfnamefont{S.~D.} \bibnamefont{Baton}},
  \bibinfo{author}{\bibfnamefont{T.}~\bibnamefont{Cowan}},
  \bibinfo{author}{\bibfnamefont{R.~R.} \bibnamefont{Freeman}},
  \bibinfo{author}{\bibfnamefont{T.}~\bibnamefont{Hall}}, \bibnamefont{et~al.},
  \bibinfo{journal}{Phys. Rev. E} \textbf{\bibinfo{volume}{69}},
  \bibinfo{pages}{066414} (\bibinfo{year}{2004}).

\bibitem[{\citenamefont{Akli et~al.}(2012)\citenamefont{Akli, Storm, McMahon,
  Jiang, Ovchinnikov, Schumacher, and Freeman}}]{Akli:PRE2012}
\bibinfo{author}{\bibfnamefont{K.~U.} \bibnamefont{Akli}},
  \bibinfo{author}{\bibfnamefont{M.~J.} \bibnamefont{Storm}},
  \bibinfo{author}{\bibfnamefont{M.}~\bibnamefont{McMahon}},
  \bibinfo{author}{\bibfnamefont{S.}~\bibnamefont{Jiang}},
  \bibinfo{author}{\bibfnamefont{V.}~\bibnamefont{Ovchinnikov}},
  \bibinfo{author}{\bibfnamefont{D.~W.} \bibnamefont{Schumacher}},
  \bibnamefont{and} \bibinfo{author}{\bibfnamefont{R.~R.}
  \bibnamefont{Freeman}}, \bibinfo{journal}{Phys. Rev. E}
  \textbf{\bibinfo{volume}{86}}, \bibinfo{pages}{026404}
  (\bibinfo{year}{2012}).

\bibitem[{\citenamefont{Welch et~al.}(2006)\citenamefont{Welch, Rose, Cuneo,
  Campbell, and Mehlhorn}}]{Welch:2006}
\bibinfo{author}{\bibfnamefont{D.~R.} \bibnamefont{Welch}},
  \bibinfo{author}{\bibfnamefont{D.~V.} \bibnamefont{Rose}},
  \bibinfo{author}{\bibfnamefont{M.~E.} \bibnamefont{Cuneo}},
  \bibinfo{author}{\bibfnamefont{R.~B.} \bibnamefont{Campbell}},
  \bibnamefont{and} \bibinfo{author}{\bibfnamefont{T.~A.}
  \bibnamefont{Mehlhorn}}, \bibinfo{journal}{Phys. Plasmas}
  \textbf{\bibinfo{volume}{13}}, \bibinfo{pages}{063105}
  (\bibinfo{year}{2006}).

\bibitem[{\citenamefont{Link et~al.}(2011)\citenamefont{Link, Freeman,
  Schumacher, and Woerkom}}]{Link:POP2011}
\bibinfo{author}{\bibfnamefont{A.}~\bibnamefont{Link}},
  \bibinfo{author}{\bibfnamefont{R.~R.} \bibnamefont{Freeman}},
  \bibinfo{author}{\bibfnamefont{D.~W.} \bibnamefont{Schumacher}},
  \bibnamefont{and} \bibinfo{author}{\bibfnamefont{L.~D.~V.}
  \bibnamefont{Woerkom}}, \bibinfo{journal}{Phys. Plasmas}
  \textbf{\bibinfo{volume}{18}}, \bibinfo{pages}{053107}
  (\bibinfo{year}{2011}).

\bibitem[{\citenamefont{Yang}(2009)}]{JHYang:Thesis}
\bibinfo{author}{\bibfnamefont{J.-H.} \bibnamefont{Yang}}, Ph.D. thesis,
  \bibinfo{school}{University of Rochester} (\bibinfo{year}{2009}).

\bibitem[{\citenamefont{Krygier}(2013)}]{Andy:Thesis}
\bibinfo{author}{\bibfnamefont{A.}~\bibnamefont{Krygier}}, Ph.D. thesis,
  \bibinfo{school}{The Ohio State University} (\bibinfo{year}{2013}).

\bibitem[{\citenamefont{Krygier et~al.}()\citenamefont{Krygier, Schumacher, and
  Freeman}}]{Andy:unpublished}
\bibinfo{author}{\bibfnamefont{A.}~\bibnamefont{Krygier}},
  \bibinfo{author}{\bibfnamefont{D.~W.} \bibnamefont{Schumacher}},
  \bibnamefont{and} \bibinfo{author}{\bibfnamefont{R.~R.}
  \bibnamefont{Freeman}}, \bibinfo{note}{(unpublished)}.

\bibitem[{\citenamefont{Sentoku et~al.}(2004)\citenamefont{Sentoku, Mima, Ruhl,
  Toyama, Kodama, and Cowan}}]{Sentoku:POP2004}
\bibinfo{author}{\bibfnamefont{Y.}~\bibnamefont{Sentoku}},
  \bibinfo{author}{\bibfnamefont{K.}~\bibnamefont{Mima}},
  \bibinfo{author}{\bibfnamefont{H.}~\bibnamefont{Ruhl}},
  \bibinfo{author}{\bibfnamefont{Y.}~\bibnamefont{Toyama}},
  \bibinfo{author}{\bibfnamefont{R.}~\bibnamefont{Kodama}}, \bibnamefont{and}
  \bibinfo{author}{\bibfnamefont{T.~E.} \bibnamefont{Cowan}},
  \bibinfo{journal}{Phy. Rev. Lett.} \textbf{\bibinfo{volume}{11}},
  \bibinfo{pages}{3083} (\bibinfo{year}{2004}).

\bibitem[{\citenamefont{Nakamura et~al.}(2004)\citenamefont{Nakamura, Kato,
  Nagatomo, and Mima}}]{Nakamura:PRL2004}
\bibinfo{author}{\bibfnamefont{T.}~\bibnamefont{Nakamura}},
  \bibinfo{author}{\bibfnamefont{S.}~\bibnamefont{Kato}},
  \bibinfo{author}{\bibfnamefont{H.}~\bibnamefont{Nagatomo}}, \bibnamefont{and}
  \bibinfo{author}{\bibfnamefont{K.}~\bibnamefont{Mima}},
  \bibinfo{journal}{Phy. Rev. Lett.} \textbf{\bibinfo{volume}{93}},
  \bibinfo{pages}{265002} (\bibinfo{year}{2004}).

\end{thebibliography}

\end{document}